\documentclass{aa}  

\usepackage{graphicx}
\usepackage{txfonts}
\usepackage{lipsum}
\usepackage{subcaption}         % necessary for continued figures, example in section 3
\usepackage{lscape}             % to rotate a single page table, example in appendix.
\usepackage{placeins}           % useful with \FloatBarrier, to keep 
\usepackage{upgreek}

\usepackage{orcidlink}

\usepackage{ulem}

\usepackage{graphicx}

\begin{document}

%%%%%%%%%%%%%%%%%%%%%%%%%%%%%%%%%%%%%%%%
% if you use custom commands in your title,
% ensure to check your title when submitting!
%%%%%%%%%%%%%%%%%%%%%%%%%%%%%%%%%%%%%%%%
   \title{Discovery of the first octupole pulsation mode in a $\delta$ Scuti star}

   \subtitle{A stationary $\ell = 3$ sectoral mode}

%%%%%%%%%%%%%%%%%%%%%%%%%%%%%%%%%%%%%%%%
% Please separate each author with the \and command
%
% Please do not include ORCIDs next to author names.
% Only ORCIDs authenticated by individual authors in EDPS
% editorial system will be taken into account.
% ORCIDs included here will be removed.
%%%%%%%%%%%%%%%%%%%%%%%%%%%%%%%%%%%%%%%%

   \author{S. A. Rappaport\orcidlink{0000-0003-3182-5569}\,\inst{1,2} 
          \and
          R. Jayaraman\orcidlink{0000-0002-7778-3117}\,\inst{3}
	  \and
	  G. Handler\orcidlink{0000-0001-7756-1568}\,\inst{4}
	  \and
	  D. Kurtz\orcidlink{0000-0002-1015-3268}\,\inst{5,6}
	  \and
	  V. Zhang\orcidlink{0009-0007-2660-7635}\,\inst{7}
	  \and
	  R. Gagliano\orcidlink{0000-0002-5665-1879}\,\inst{8} 
	  \and
	  B. Powell\orcidlink{0000-0003-0501-2636}\,\inst{9}
	  \and
	  J. Fuller\orcidlink{0000-0002-4544-0750}\,\inst{10}
	  \and
	  T. Borkovits\orcidlink{0000-0002-8806-496X}\,\inst{1,11,12,13,14} 
	  \and
	  V. Kostov\orcidlink{0000-0001-9786-1031}\,\inst{9,15}
	  \and
	  J. Daszy\'{n}ska-Daszkiewicz\orcidlink{0000-0001-9704-6408}\,\inst{16,17}
         }

   \institute{HUN--REN--SZTE Stellar Astrophysics Research Group, H-6500 Baja, Szegedi \'ut, Kt. 766, Hungary %1
	\and
	  Department of Physics, Kavli Institute for Astrophysics and Space Research, M.I.T., Cambridge, MA 02139, USA \\ %2
	   \email{sar@mit.edu}
         \and
         Department of Astronomy, Cornell University, 122 Sciences Dr, Ithaca, NY 14850 USA  %3
         \and
         Nicolaus Copernicus Astronomical Center, Polish Academy of Sciences, ul. Bartycka 18, PL-00-716 Warszawa, Poland %4
         \and
         Centre for Space Research, North-West University, Mahikeng 2745, South Africa %5
         \and
         Jeremiah Horrocks Institute, University of Lancashire, Preston PR1 2HE, UK   %6
         \and
         Harvard University, Cambridge, MA 02138, USA %7
         \and
         Amateur Astronomer, Glendale, AZ 85308  %8
	\and
	   NASA Goddard Space Flight Center, 8800 Greenbelt Road, Greenbelt, MD 20771, USA  %9
	\and
	TAPIR, Mailcode 350-17, California Institute of Technology, Pasadena, CA 91125, USA %10
	\and
   	   Baja Astronomical Observatory of University of Szeged, H-6500 Baja, Szegedi \'ut, Kt. 766, Hungary.\\
             \email{borko@electra.bajaobs.hu}  %11
	\and
	   Konkoly Observatory, Research Centre for Astronomy and Earth Sciences,  H-1121 Budapest, Konkoly Thege Mikl\'os \'ut 15-17, Hungary   %12
	\and
	   ELTE E{\"o}tv{\"o}s Lor\'and University, Gothard Astrophysical Observatory, Szent Imre h. u. 112, 9700 Szombathely, Hungary %13
	\and
	   HUN--REN--ELTE Exoplanet Research Group, H-9700 Szombathely, Szent Imre h. u. 112, Hungary %14
	\and
	   SETI Institute, 189 Bernardo Avenue, Suite 200, Mountain View, CA 94043, USA %15
        \and
	   Astronomical Institute of the Wroc\l aw University, ul. Kopernika 11, 51-622 Wroc\l aw, Poland %16
        \and
	Copernicus Astronomical Center, Bartycka 18, 00-716 Warsaw, Poland %17
             }
   \date{Received \today}

% \abstract{}{}{}{}{}
% 5 {} token are mandatory
 
  \abstract
   {}
   {We are attempting to better understand how stellar pulsations in close binary systems are affected, and possibly induced, by tidal, Coriolis, and centrifugal forces.} 
   {We analyzed TESS data for some 50,000 potential eclipsing binaries selected by machine learning algorithms in order to search for pulsation multiplets split by integer multiples of the orbital frequency.}
   {We report on the discovery of an octupole pulsation mode in the binary star system TIC 287869463, which contains a $\delta$\,Scuti star. This mode is actually a combination of $Y_{3+3}$ and $Y_{3-3}$ modes that are perturbed into a new eigenmode of the star via tidal, Coriolis, and centrifugal forces, which we call a `$Y_{33+}$' mode.  The mode is stationary on the star as opposed to being a traveling wave around the pulsation equator.  To our knowledge, this is the first time that such an $\ell=3$ mode identification has been securely made in any $\delta$ Scuti star, and the first stationary $\ell=3$ sectoral mode of this type seen in any star, including the Sun. The $\ell=3$ pulsations appear as a combination of two components at 34.94617 d$^{-1}$ and 39.31127 d$^{-1}$, split by exactly six times the frequency of the orbital motion to within better than 1 part in $10^5$. We extract the pulsation frequencies from the TESS data spanning more than three years, and model the system to gain a better understanding of this novel asteroseismic discovery. The pulsation frequencies are found to be steadily increasing with time, but always maintaining a split equal to six times the orbital frequency.}
   {We discuss the implications for the broader class of ``tidally tilted pulsators'' and ``tri-axial pulsators'' that have been discovered to date.  We conclude that these previous categories can all be interpreted as linear combinations of spherical harmonics whose axes coincide with the orbital axis and form new eigenmodes of the star via tidal, Coriolis, and centrifugal perturbations}


   \maketitle
   
   \nolinenumbers

%%%%%%%%%%%%%%%%%%%%%%%%%%%%%%%%%%%%%%%%%%%%%%%%%%%%%%%%%%%%%%
\section{Introduction} 

Stellar pulsations have been studied for more than a century since 1919, when Eddington began an exploration of the theory of stellar pulsation with particular interest in understanding Cepheid variables (\citealt{1919MNRAS..79..171E,1919MNRAS..79Q.177E,1926ics..book.....E}). For the first half-century of the study of stellar pulsation it was assumed that the surface geometry of pulsation modes could be described adequately by spherical harmonics \citep{cowling41}, which are the eigenmode solutions to pulsation equations for perfectly spherical stars. The eigenmodes associated with the spherical harmonics have three quantum numbers describing their geometry: $n$, the radial overtone, which is the number of radial nodes that are spherical shells, $\ell$, the degree, which is the number of surface nodes of the mode, and $m$, the azimuthal order, which gives the number of surface nodes that are lines of longitude.  It was further assumed that the axis of pulsation coincided with the rotation axis of all pulsating stars, since that, in most cases, is the principal axis of distortion from spherical symmetry (see, e.g., \citealt{kurtz22} and \citealt{2010aste.book.....A} for reviews). 

Traditionally, in stars other than the Sun, the degree of the mode is inferred, rather than observed directly. Surface cancellation of observed sectors of the spherical harmonics has meant that modes of higher degree ($\ell \ge 3$) have not been detected at all, since these higher degree modes have very low visibility (\citealt{1977AcA....27..203D}). On the other hand, because the Sun has a well-resolved surface, modes of very high degree can be detected to beyond $\ell = 300$ (\citealt{larson11}, \citealt{scherrer12}, \citealt{larson15}). Observation and knowledge of these high-degree modes place major constraints on the structure of the Sun via helioseismology.

Since we know that the Sun pulsates in very high degree modes, we can reasonably conjecture that other stars do, too. How do such high-degree modes affect the structure of a pulsating star, its evolution, its internal rotation and angular momentum transfer? The answers to these important questions can be illuminated by the direct observation of modes of degree $\ell \ge 3$. 

\citet{1982MNRAS.200..807K} discovered the rapidly oscillating Ap (roAp) stars and demonstrated for the first time that some stars have pulsation axes that are inclined to the rotation axis of the star. In the case of the strongly magnetic roAp stars, that pulsation axis is close to the magnetic axis, which is itself inclined to the rotation axis. \citet{1982MNRAS.200..807K} proposed the oblique pulsator model to explain the pulsation amplitude and phase variations seen to occur in  roAp stars with rotation. Simply, this shows that as an obliquely pulsating star rotates, the geometry of the degree of the pulsation mode becomes visible because the mode is seen from differing aspect. In effect, observers get to `walk around' the star and see the non-radial modes from different aspect, thus exposing the number and positions of the surface nodes. 

More recently, data from the Transiting Exoplanet Survey Satellite (TESS; \citealt{ricker15}) enabled the identification of two $\delta$ Scuti stars in binaries with pulsations largely confined to one hemisphere with respect to the tidal axes on the L1 side (\citealt{handler20}; \citealt{kurtz20}).  These were dubbed ``single-sided pulsators’’.  HD 74423 had one prominent pulsation mode with a half-dozen components separated by the orbital frequency $\nu_{\rm orb}$ and a systematic change in pulse phase by $\uppi$ radians every orbit.  CO\,Cam had four such modes, each with components also separated by $\nu_{\rm orb}$.  In these systems, there were systematic phase changes when the pulsation amplitude was near zero in each of these modes.  

These modes were discovered largely because of their visually striking ``single-sidedness’’, but this complication also made it difficult to understand the nature of the underlying modes themselves from the Fourier transforms (FTs).  At the time, it was assumed that the axes of these pulsations lie along the tidal axis, i.e., that they were ``tidally tilted modes’', wherein the pulsations were tidally trapped on one hemisphere and suppressed on the other \citep{fuller20}.

\citet{rappaport21} reported a much simpler multiplet mode in a $\delta$ Scuti pulsator in a binary system, with no single-sidedness.  This mode had two prominent equal-amplitude components separated by 2 $\nu_{\rm orb}$ with two weak peaks on either side of the larger peaks and separated by 1 $\nu_{\rm orb}$.  There were two maxima in this pulsation mode around the orbit (at the ellipsoidal light variation, `ELV', maxima), and phase jumps of $\uppi$ rad at the eclipses.  This was interpreted as a tidally tilted $Y_{11x}$ mode, where the x refers to the pulsation axis lying along the tidal axis (the $\hat{x}$ direction)\footnote{We take the $\hat{z}$ and $\hat{y}$ directions to lie along the angular momentum vector of the binary and the direction in the orbital plane perpendicular to the tidal axis, respectively.}.

Building on these discoveries, \citet{jayaraman22} identified 31 pulsation modes a pulsating sdB star with a white dwarf companion. These modes typically had between 3 and 5 components separated by $\nu_{\rm orb}$, with one or two $\uppi$ phase jumps per orbit. These were not simple to understand, but were interpreted at the time as tidally tilted $Y_{{\rm 1m}x}$ or $Y_{{\rm 2m}x}$ modes.

More recently, TESS data revealed two $\delta$ Scuti stars in binary systems with rich pulsation spectra---9 and 14 doublets (\citealt{zhang24}; \citealt{jayaraman24}), which we interpreted as simple dipole pulsation modes.  Each of these modes consists of just two dominant components separated by 2 $\nu_{\rm orb}$ with either no detectable central component, or a weak one.  Each mode has two amplitude maxima per orbit, either at the eclipses, or at the ELV maxima.  Each also has two $\uppi$ phase jumps per orbit, either at the ELV maxima, or at the eclipses, respectively.  These were interpreted as $Y_{\rm 10x}$ or $Y_{\rm 10y}$ modes, depending on where the modes had their amplitude maxima.  

\citet{fuller25} modeled these systems as an entirely new type of mode which consist of linear combinations of $Y_{1+1z}$ and $Y_{1-1z}$ modes that are coupled by tidal, centrifugal, and Coriolis forces in the binary.  This modeling revealed that linear combinations of $Y_{1+1z}$ and $Y_{1-1z}$ modes produce, and are mathematically identical to, tidally tilted modes $Y_{10x}$ and $Y_{10y}$.  Thus, either interpretation is valid.  However, as \citet{fuller25} also showed, other higher order modes with pulsation axes along $z$ (i.e., with $\ell =2$ and $\ell =3$) can also be combined via tidal, centrifugal, and Coriolis forces to produce new eigenmodes of the star. These modes also have multiplet  frequencies that are spaced by integer multiples of $\nu_{\rm orb}$ and can exhibit up to 4 and 6 amplitude maxima, respectively, per orbit as well as up to 4 and 6\,$\uppi$ phase jumps.  

These latter Fuller modes are all linear combinations of spherical harmonics with pulsation axes lying along the spin (or orbital) axis. To the lowest order, modes of the same $\ell$ and differing by $\Delta m = 2$ are coupled by the tidal forces \citep{fuller25}.   We note that for $\ell \ge 2$ the degeneracy is broken, and these are no longer equal to, or even mimic, simple pulsation modes that have been tidally tilted along either the x or y axes.  Thus, it is now possible within this paradigm to distinguish between  tidally coupled ``$z$'' modes and tidally tilted modes whose pulsation axes lie along $\hat{x}$ or $\hat{y}$.

All of the `tidally tilted' and `tri-axial' modes mentioned above can be reinterpreted as  Fuller modes, mostly with $\ell = 1$ and $\ell =2$ components that may or may not be substantially suppressed in one of the tidal hemispheres.  Here, we report the discovery of a $\delta$ Scuti pulsator in a binary system (TIC 287869463) which exhibits a Fuller octupole pulsation mode that cannot be described as any kind of tidally tilted pulsation. While $\ell = 3$ pulsation modes have been found in red giants (\citealt{stello16}), as well as in some main sequence solar-like oscillators (e.g., \citealt{kjeldsen05}), and suggested in some $\beta$ Cephei stars (e.g., \citealt{aerts98}, \citealt{cotton22}), no such modes have been robustly detected in $\delta$ Scuti stars.  Moreover, no previously-discovered $\ell = 3$ sectoral mode is stationary, i.e., non-circulating. 

In Sect.~\ref{sec:discovery} we present observational evidence for these modes and show how the amplitudes of these modes vary with time, while the phase differences between the upper and lower frequency components remain constant to within the statistical uncertainties.  Simulations of these and other Fuller modes are presented in Sect.~\ref{sec:simulations}, wherein we highlight the striking similarity between the simulated frequency peaks and what we observe in TESS data.  An analysis of the binary system parameters is undertaken in Sect.~\ref{sec:system}.  We summarize and discuss our results in Sec.~\ref{sec:discussion}.  In the Appendix, we discuss O-C diagrams for the pulsations and show that the pulsation frequencies are steadily increasing over the 5-yr interval of the TESS observations.

\begin{table}
\centering
\caption{Sectors and cadence for TIC 287869463}
\small
\begin{tabular}{lc}
\hline
\hline
S87, S90, S93 & 200-s \\
S63, S64, S65, S66 & 200-s  \\ 
S30, S33, S36, S37, S39 & 600-s  \\
S3, S6, S10, S11, S12, S13 & 30-min \\
\hline
\label{tbl:sectors}  % Table 1
\end{tabular}
%\textit{Notes.}  
\end{table}

\begin{table}
\centering
\caption{Properties of TIC 287869463$^a$}
\small
\begin{tabular}{lc}
\hline
\hline
RA [degree] & 127.9633 \\
Dec [degree] &  $-75.4937$  \\ 
$G$     &  11.87  \\
$B_p$  & 12.01 \\
$R_p$  &  11.63 \\
$B_p-R_p$ & 0.379 \\
pmra [mas yr$^{-1}$]  & 1.89 \\
pmdec [mas yr$^{-1}$]  & $-2.42$ \\
radial velocity [km s$^{-1}$] & 27.58 \\
RV$\_$amp$\_$robust$^b$ [km s$^{-1}$] & 144 \\
distance [pc]$^c$ & $1086 \pm 3$ \\
$P_{\rm orb}$ [d]$^d$ & 1.374\,536 \\
\hline
\label{tbl:properties}  % Table 2
\end{tabular}

\textit{Notes.}  (a) From Gaia DR3 \citep{GaiaEDR3} unless otherwise noted. (b) This parameter, taken from Gaia, represents approximately twice the $K_1$ amplitude \citep{katz23}. (c) \citet{bailer-jones21}. (d) From this work. 
 
\end{table} 
%%%%%%%%%%%%%%%%%%%%%%%%%%%%%%%%%%%%%%%%%%%%%%%%%%%%%%%%%%%%%%

\section{Discovery of Fuller modes in TIC 287869463} \label{sec:style}
\label{sec:discovery}

We are currently carrying out an extensive search for Fuller-mode pulsators in TESS data. In particular, we are examining a set of bright likely eclipsing binaries (TESS magnitude $T_{\rm mag} < 13.5$) selected by machine learning from 56  sectors of TESS observations with 10-minute cadence or shorter (\citealt{kostov25}; \citealt{powell26}; Jayaraman et al., in prep).  In all, there are 51,820 such candidate binaries whose $T_{\rm eff}$ falls in the range for them to contain possible $\delta$ Scuti pulsators (6500 K $\leq T_{\rm eff} \leq 9000$ K). The TESS light curves used for the search are taken from the MIT Quick-Look Pipeline (QLP; \citealt{kunimoto21,kunimoto22}).  

For each candidate binary, we produce an automated echelle diagram, which plots the pulsation frequency against the ``echelle phase'' (the pulsation frequency $\nu_{\rm puls}$ modulo $\nu_{\rm orb}$, normalized to $\nu_{\rm orb}$). To do so, we first determined the orbital period $P_{\rm orb}$, if any, and then subtracted the first 30 harmonics of $\nu_{\rm orb}$ from the light curve. We then sequentially identified and removed from the remaining light curve the 75 highest-amplitude pulsation frequencies in the FT, or the highest-amplitude pulsations down to a noise floor of 5 times the rms amplitude in the FT---whichever came first. The echelle diagrams were created from these highest-amplitude peaks. Finally, we examined each echelle diagram by eye for interesting modes that are split by integer multiples of the orbital frequency. Most of the split modes are dipoles (split by 2$\nu_{\rm orb}$), along with a smaller percentage of quadrupoles (split by 4 $\nu_{\rm orb}$). However, we identified one target (TIC 287869463 or Gaia DR3 5216608316411701248; \citealt{stassun19}) as having an octupole mode, split by 6\,$\nu_{\rm orb}$.

\begin{figure}[h]
\centering
\includegraphics[width=0.46\textwidth]{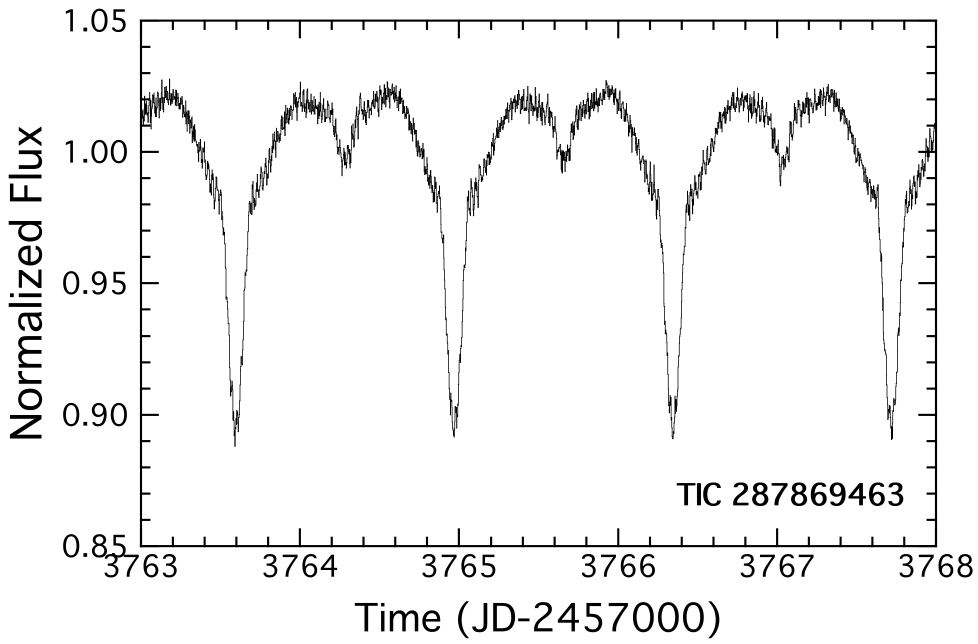}
\includegraphics[width=0.45\textwidth]{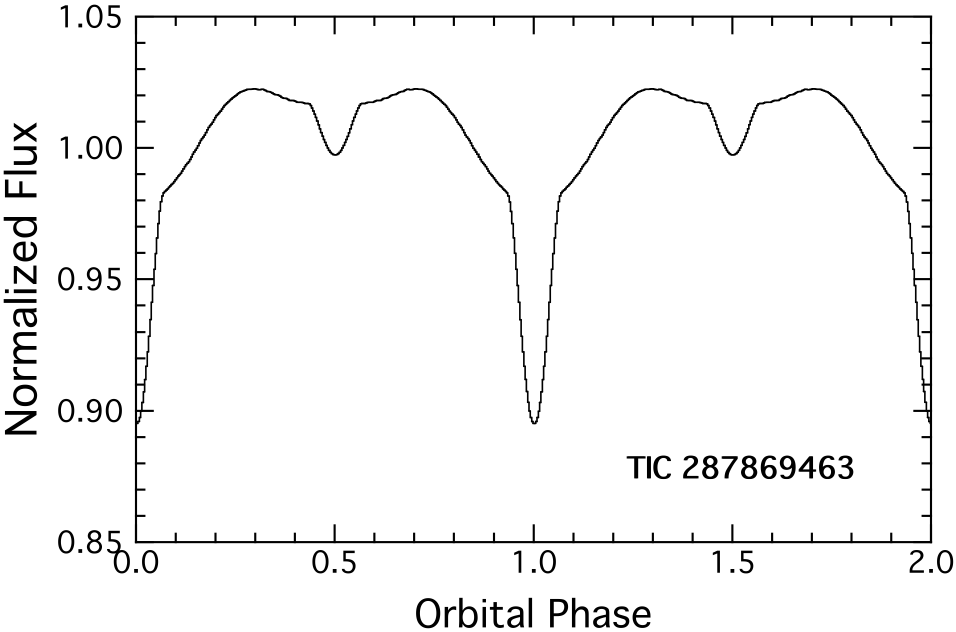}
\caption{Light curves for TIC 287869463. \textit{Top}: 5-d segment of the raw TESS light curve. The pulsations superposed on the eclipsing light curve are readily apparent. \textit{Bottom}: Fourier-reconstructed light curve from the first 60 orbital harmonics. Here, we used only the cosine terms in the reconstruction to remove a small, time-varying, O'Connell effect \citep{oconnell51}, presumably arising from star spots on the cooler companion star. }
\label{fig:lcs}
\end{figure}  % Figure 1

A short 5-d segment of the TESS light curve for TIC\,287869463 is shown in Figure~\ref{fig:lcs} (top panel).  The stellar pulsations of the hotter component can be seen superposed on the eclipsing light curve.  In all, there are 7 sectors of data with a 200-s cadence, for a total of about 168\,d of data, with 100\,d of that being contiguous. A Fourier-reconstructed light curve made from the first 60 orbital harmonics is given  in Figure~\ref{fig:lcs} (bottom panel). Table \ref{tbl:sectors} lists all TESS sectors in which TIC 287869463 was observed, alongside the observing cadence. 
Table \ref{tbl:properties} enumerates photometric and astrometric information about TIC 287869463, and the
system properties (modeled in Section \ref{sec:system}).

\begin{figure}
\centering
\includegraphics[width=0.48\textwidth]{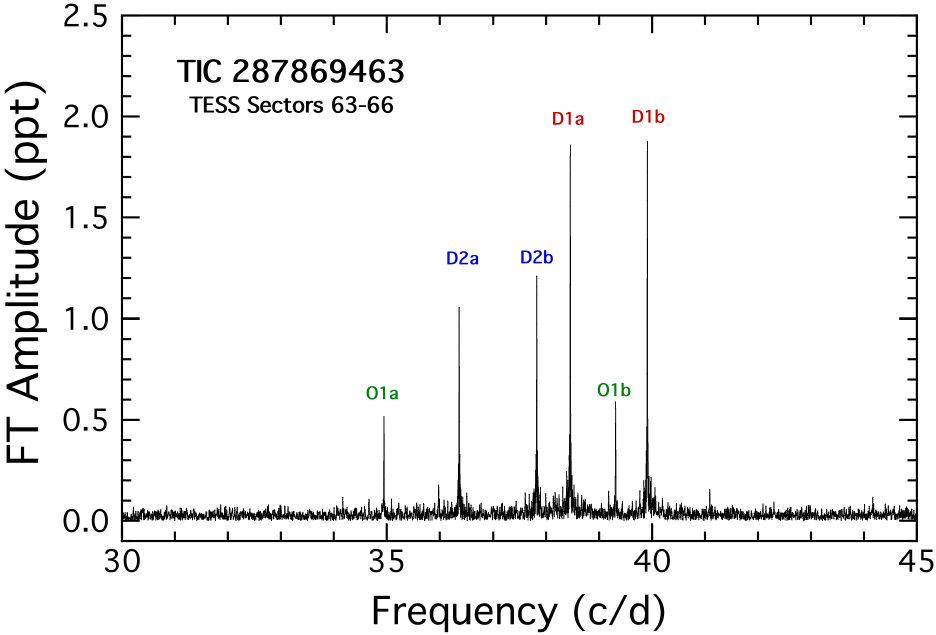}
\caption{Fourier transform of the TESS data from Sectors 63 to 66 for TIC 28786963.  The labels mark the two dipole modes (`D1' and `D2') and an octupole mode (`O1'). Each mode has two prominent components, marked `a' and `b'.  The dipole components are separated by 2 $\nu_{\rm orb}$, while the octupole components are separated by 6 $\nu_{\rm orb}$.}
\label{fig:ft}
\end{figure}  % Figure 2

\begin{figure}[h]
\centering
\includegraphics[width=0.9\columnwidth]{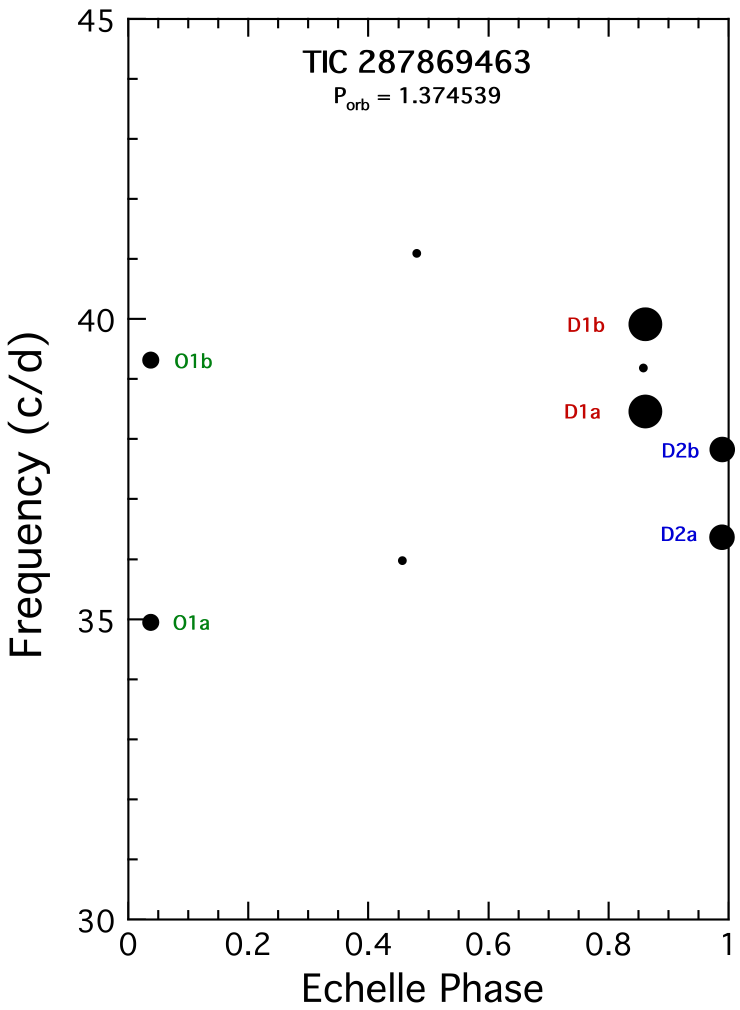}
\caption{Echelle diagram for TIC 287869463 constructed from the TESS data of Sectors 63-66. This shows the frequencies of the pulsations vs echelle phase---the frequency modulo $\nu_{\rm orb}$ and normalized to  $\nu_{\rm orb}$. The labeling of the components follows Fig.~\ref{fig:ft}. }
\label{fig:echelle}
\end{figure}  % Figure 3

\begin{figure}
\centering
\includegraphics[width=0.48\textwidth]{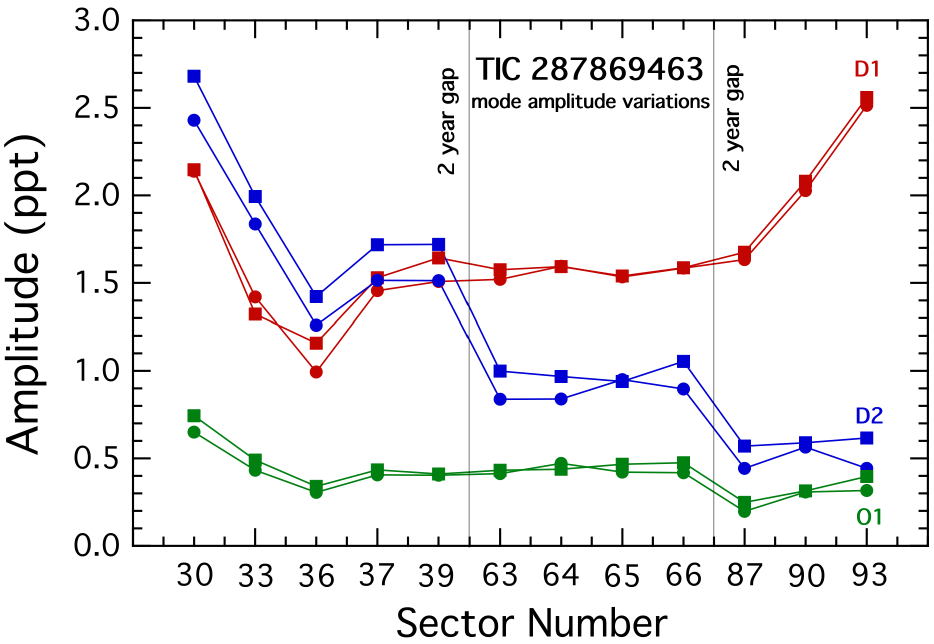}   \hglue0.1cm 
\includegraphics[width=0.48\textwidth]{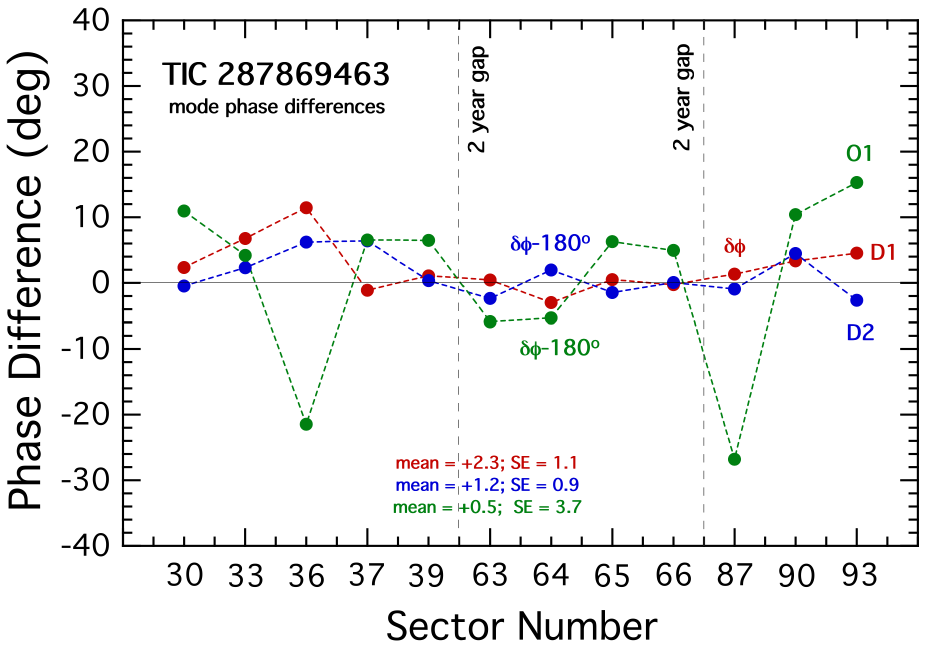}
\caption{\textit{Top}: Variations in the dipole (red and blue) and octupole (green) component amplitudes with sector number across $\sim$5\,yr. Note that for each mode, the two component peaks have very similar amplitudes, and their variations are clearly tightly correlated. \textit{Bottom}: The difference in phase between the two components of each of the three modes, at the times of primary eclipse. To fit all three curves on the same scale, we subtracted 180 degrees from the phase difference of modes D2 and O1. This shows that the phase differences remain constant at either 0 or $\uppi$ rad to within the few degrees of our statistical precision.  This analysis shows that the geometry of the  modes remains unchanged over the time span of the observations. 
}
\label{fig:amp_phase}
\end{figure}  % Figure 4

The Fourier transform of the 200-s cadence data from Sectors 63-66 (100\,d of nearly continuous data) is shown in Fig.~\ref{fig:ft}. The data were cleaned of the first 40 orbital harmonics before computing the FT. The FT shows 6 prominent peaks due to stellar pulsations, all in the frequency range $\sim$35-40 d$^{-1}$. The dipole pulsation modes (separated by 2$\nu_{\rm orb}$) are labeled ``D1'' and ``D2'' , and the octupole mode---separated by 6$\nu_{\rm orb}$---is labeled as ``O1.'' 

The echelle diagram for this star's pulsations is shown in Figure ~\ref{fig:echelle}.  The D1 mode has a small central component, while D2 does not.  The octupole mode is split by 6 $\nu_{\rm orb}$, and has only one tiny intermediate frequency peak, which is situated at 1 $\nu_{\rm orb}$ from the missing central frequency.  There are also two low amplitude singlet modes at 35.976 and 41.096 d$^{-1}$.  

\begin{table*}
\centering
\caption{Frequencies of the 6 prominent pulsations seen in TIC 287869463$^a$}
\small
\begin{tabular}{lccccc}
\hline
\multicolumn{1}{c}{Frequency} & \multicolumn{1}{c}{Amplitude} & \multicolumn{1}{c}{Phase Difference$^b$} & \multicolumn{1}{c}{Mode ID} & \multicolumn{1}{c}{Frequency Split} & \multicolumn{1}{c}{Inferred $P_{\rm orb}$$^c$} \\
 \multicolumn{1}{c}{d$^{-1}$} & \multicolumn{1}{c}{mmag} & \multicolumn{1}{c}{degrees} & \multicolumn{1}{c}{...}& \multicolumn{1}{c}{d$^{-1}$} & \multicolumn{1}{c}{d} \\
\hline
38.45620(1)   & $1.0-2.5$ & $+2.3 \pm 1.1$ & Dipole 1a & ... & ... \\  
39.91123(1)   &  $1.2-2.6$  & $+2.3 \pm 1.1$ & Dipole 1b & $2 \times 0.727516(20)$ & 1.374539 \\ 
36.36627(2)   &  $0.5-2.4$ & $181.2 \pm 0.9$  & Dipole 2a & ... & ... \\  
37.82130(2)  &   $0.6-2.7$ & $181.2 \pm 0.9$ & Dipole 2b & $2 \times 0.727514(34)$ & 1.374544  \\ 
34.94617(3)  &   $0.2-0.6$ & $180.4 \pm 3.7$ & Octupole 1a & ... & ... \\ 
39.31127(3)  &   $0.3-0.7$ & $180.4 \pm 3.7$ & Octupole 1b & $6 \times 0.727518(39)$ & 1.374537 \\ 
\hline
\label{tbl:freqs}  % Table 3
\end{tabular}

\textit{Notes.}  (a) All frequencies are referenced to an epoch of BJD 2460000.  Values in parentheses are the uncertainties in the last digits.  See Table \ref{tbl:freq_summary} for details of how the frequencies are changing with time.  (b) Average phase differences between the two mode components.  See Fig.~\ref{fig:amp_phase} for a graphical representation.  Phases are taken at the time of primary eclipse. (c) The orbital period at the common epoch is 1.374,536,0(2) d.
\end{table*} 

The six most significant frequencies seen in the periodogram and echelle diagram, related to D1, D2, and O1, are enumerated in Table \ref{tbl:freqs}. These frequencies were produced from phase tracking of the pulsations (see, e.g., \citealt{bowman16}, and also Appendix \ref{sec:phase_tracking}) over all 12 sectors of 200-s and 600-s cadence data (where the pulsation frequencies were below the Nyquist frequency). All frequencies are referenced to a common epoch, BJD = 2460000. In the 5th column of Table \ref{tbl:freqs}, we list the mean frequency split between the two components of the particular mode.  In the last column, we indicate the (orbital) period inferred from the splitting, and these values all agree to within $10^{-5}$ d of the calculated orbital period\footnote{The apparent orbital period is also changing with time, and the cited orbital period is also referenced to this same epoch of BJD = 2460000 (see Appendix \ref{sec:phase_tracking}).}.

The amplitudes and phases of the 6 pulsations as a function of TESS sector are given in the top and bottom panels of Fig.~\ref{fig:amp_phase}, respectively. The amplitudes of the two components of each of the dipoles are clearly correlated,
as are the two amplitudes of the octupole mode. However, the latter are less variable than for the D1 and D2 modes. The two amplitudes for each pulsation mode are always very similar and are not varying independently.  

In the phase plot (Fig.~\ref{fig:amp_phase}, bottom panel), we show the phase difference between the two components of each mode as a function of TESS sector, all referenced to a time of primary eclipse (BJD = 2460014.4757). For the D1 mode, the two components are mostly always in phase at the primary eclipse.  The two components of both the D2 and O1 modes are essentially always out of phase by 180$^\circ$ at the primary eclipses.  (This in turn means that these modes are at an amplitude minimum at primary eclipse.) The standard errors in the phase differences are only 1.1$^\circ$, $0.9^\circ$, and $3.7^\circ$, for the D1, D2, and O1 modes, respectively.

\begin{figure}
\centering
\includegraphics[width=0.98\columnwidth]{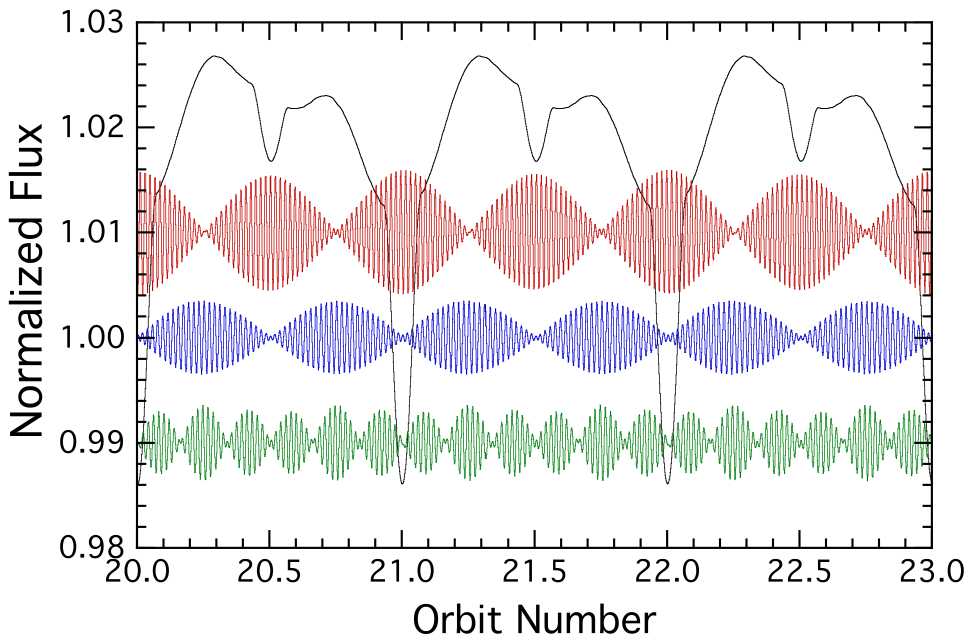} \hglue0.1cm
\includegraphics[width=0.96\columnwidth]{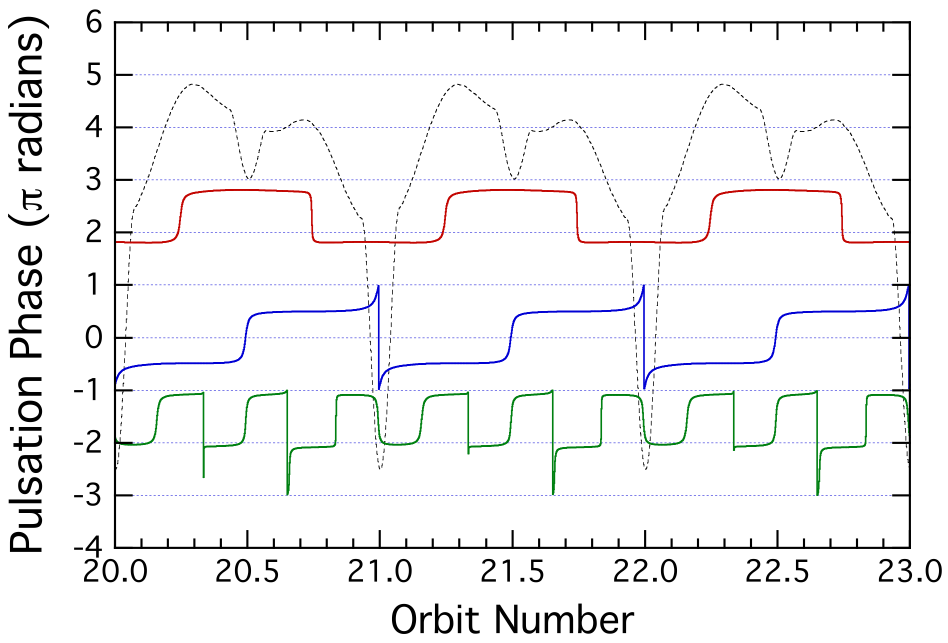}
\caption{Reconstructions of pulsation amplitude and phase vs~the orbital phase. The color coding, top to bottom, is red = D1, blue = D2, and green = O1. Superposed is a scaled version of the orbital light curve to guide the eye on the orbital phasing. Note the 6 maxima in pulsation amplitude and six $\uppi$-rad phase jumps per orbit for the octupole (O1). }
\label{fig:reconstr}
\end{figure}  % Figure 5

We next force fit a three-element triplet in frequency (separated by $ \nu_{\rm orb}$) to each of the dipole modes, and a 7-element septuplet to the octupole mode.  We use the amplitudes and phases of the extracted mode elements to reconstruct the amplitudes of the three pulsation modes as a function of orbital phase. These results are shown in the top panel of Fig.~\ref{fig:reconstr}. An amplitude-scaled orbital light curve is shown superposed for phasing reference. Each of the dipoles has two amplitude maxima per orbit, while the octupole has six such maxima. We find that D1 has a maximum amplitude at the eclipses, while D2 has a minimum there.  The octupole mode also has a pulsation minimum at the eclipses. These are all in accord with the discussion above of the phase differences of the two elements of each mode (see also Fig.~\ref{fig:amp_phase}) at the time of primary eclipse. The bottom panel of Fig.~\ref{fig:reconstr} shows the run of reconstructed pulse phase vs~orbital phase.  The dipole modes have two $\uppi$ phase jumps each orbit, while the octupole mode has six $\uppi$ phase jumps. 

 There are several arguments that point to the octupole mode in this star---the first such stationary $\ell=3$ sectoral mode discovered---being a single pulsation mode split by 6 $\nu_{\rm orb}$, as opposed to two independent modes that are coincidentally separated by $\sim$6 $\nu_{\rm orb}$.  First, the separation of the octupole mode elements corresponds to an integer multiple of the measured orbital period to one part in $10^5$.  The odds of this occurring at random for the two significant non-dipole frequencies over the pulsation range of 30-40 d$^{-1}$ is only $1.7 \times 10^{-6}$ (verified by Monte Carlo simulations). Second, the very similar amplitudes of the two mode components, and their correlated amplitude and frequency variations over a 5-year interval (see Appendix \ref{sec:phase_tracking}), also strongly suggest that these are not from independent modes. Third, the fact that this mode has either a minimum or maximum at the time of primary eclipse (minimum in this case) would only occur at random about 25\% of the time.\footnote{This results from the fact that in such an octupole mode, there will be either a minimum or maximum every 30$^\circ$, and we have an accurate determination of the mode phase to within $\pm 3.7^\circ$} We also note that there is a well-developed model \citep{fuller25} that predicts such modes, and this adds confidence to the interpretation of this as an octupole mode.

%%%%%%%%%%%%%%%%%%%%%%%%%%%%%%%%%%%%%%%%%%%%%%%%%%%%%%%%%%%%%%
\section{Simulations of the modes and interpretation}
\label{sec:simulations}

\begin{figure*}
\sidecaption
\includegraphics[width=12cm]{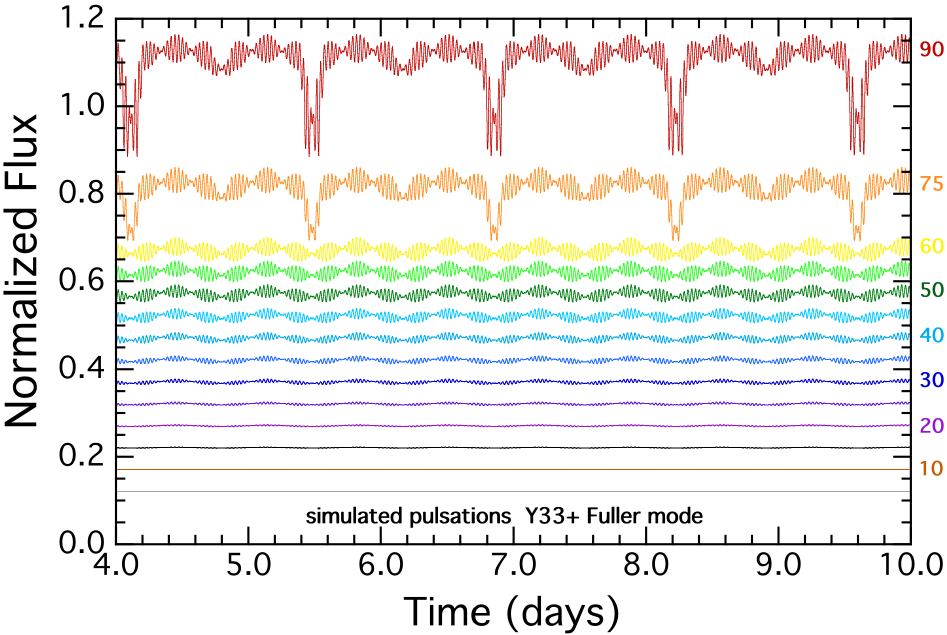}
\caption{Simulated light curves of TIC 287869463 with a $Y_{33+}$ pulsation mode.  The orbital inclination angles in degrees are written next to the right-hand $y$ axis. The orange curve for $i = 75^\circ$ represents the simulated light curve for the approximate inclination angle of TIC 287869463. The curves for different inclinations are shifted vertically by arbitrary amounts for clarity.}
\label{fig:sim_LCs}
\end{figure*}  % Figure 6

Our analysis of the TESS data for this star has shown that we have found an octupole mode of the form:
\begin{eqnarray}
\label{eqn:y33+}
Y_{33+} & = & \left(Y_{3+3z} + Y_{3-3z}\right)e^{i \omega t} \\   \nonumber  
              & = & \left(y^3 - 3 x^2 y \right) \sin(\omega t),
\end{eqnarray}
where the $Y$'s on the right hand side of the top equation are ordinary spherical harmonics
with $\ell = 3$ and $m = \pm 3$.  The extra `$z$' designation indicates that the axis of these modes lies along the rotation vector of the pulsating star, which in turn we assume is aligned with the orbital angular momentum vector (along $\hat{z}$). These spherical harmonics each have a traveling wave propagating around the star's equator.  \citet{fuller25} showed how such modes can be coupled by perturbations on the pulsating star in a binary from tidal, Coriolis, and centrifugal forces into a stationary mode (standing wave) in the rotating star's reference frame.

The $Y_{33+}$ mode is a standing wave sectoral octupole mode in a frame rotating with the orbital motion. Here, the + sign indicates the sum, rather than the difference, between the terms on the right side of Equation \ref{eqn:y33+}. This mode differs from standard sectoral $Y_{3+3z}$ or $Y_{3-3z}$ modes due to its stationary nature. Throughout the orbital cycle, we can therefore observe this mode from varying angles. This gives the same advantage of providing information on the geometry of the mode (and its viewing aspect), as is gained from oblique pulsation in roAp stars. Such geometric information cannot be extracted from rotationally perturbed normal sectoral octupole modes pulsating about the rotation axis.

\begin{figure*}
\centering
\includegraphics[width=0.48\textwidth]{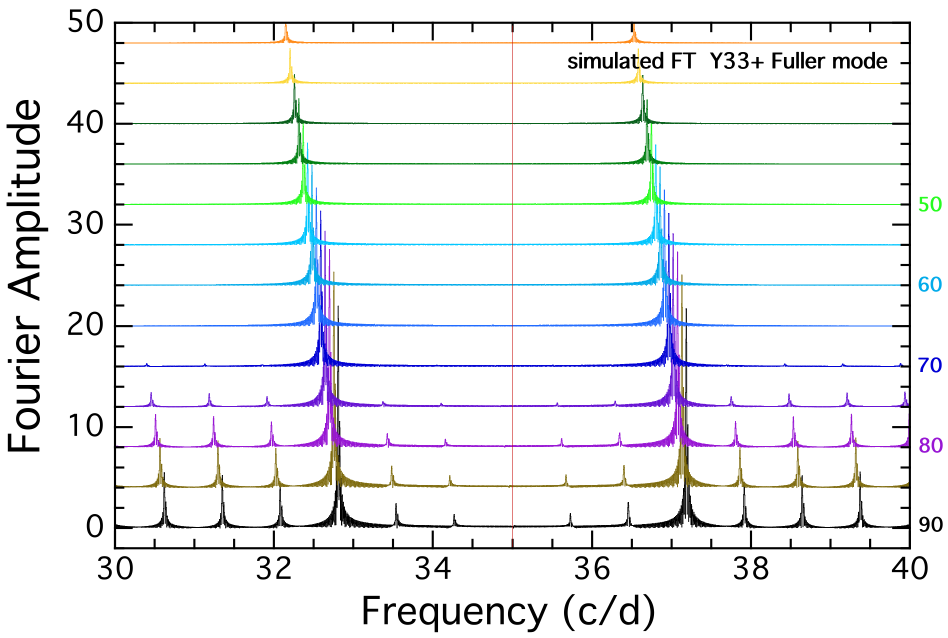}
\includegraphics[width=0.48\textwidth]{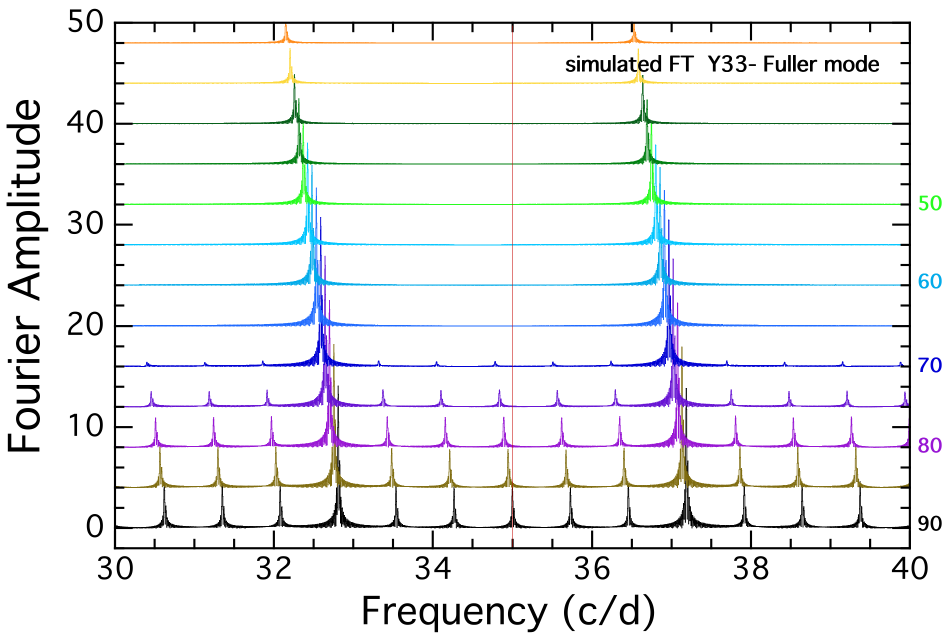}
\includegraphics[width=0.48\textwidth]{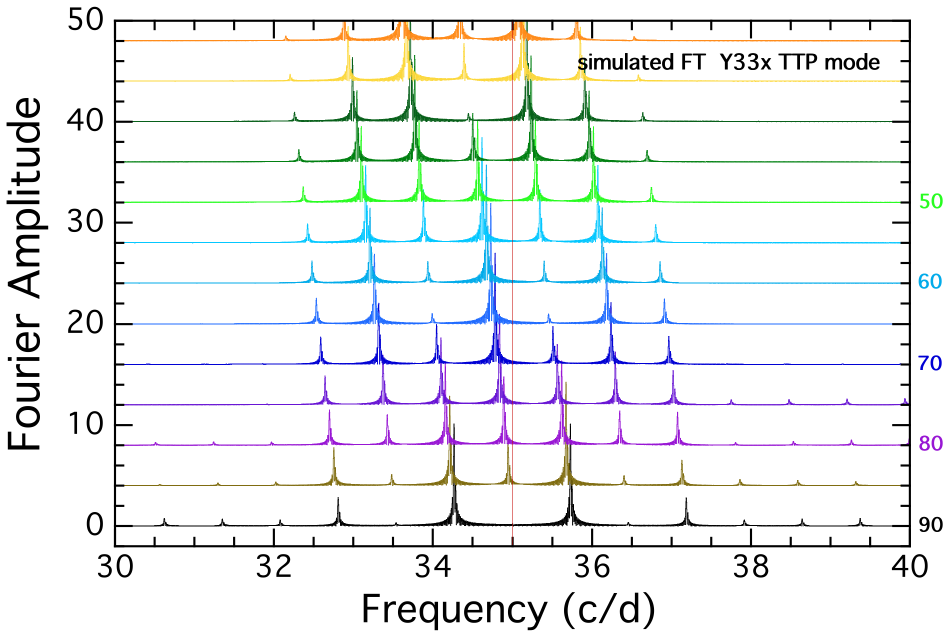}
\includegraphics[width=0.48\textwidth]{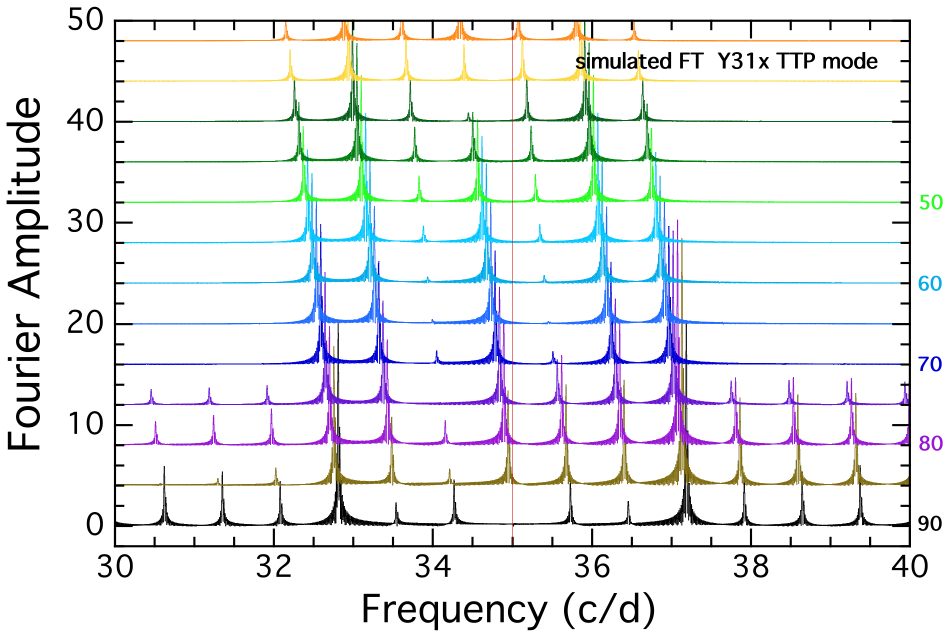}
\caption{Simulated Fourier transforms for different examples of $\ell =3$ pulsation modes. The modes represented in the four panels are $Y_{33+}$, $Y_{33-}$, $Y_{33x}$, and $Y_{31x}$, clockwise starting from the upper left panel. The first two modes are defined in Sect.~\ref{sec:simulations} and equation (\ref{eqn:y33+}).  The latter two modes have had their pulsation axis tilted into the orbital plane, and lying along the tidal (or `$x$') axis. The FT peaks are arbitrarily shifted to the right for decreasing inclination angles to avoid overlapping peaks.}
\label{fig:sim_FTs}
\end{figure*}  % Figure 7

Previously, we identified numerous stationary dipole modes formed from $Y_{11\pm}$ modes ($Y_{1+1z} \pm Y_{1-1z}$) \citep{zhang24,jayaraman24}.  We are also working on a smaller set of other pulsators with related quadrupole modes formed as $Y_{22\pm} = Y_{2+2z} \pm Y_{2-2z}$ (see, e.g., \citealt{handler25}).  As described in \citet{fuller25}, there are three other perturbed stationary $\ell = 2$ modes, and five other $\ell = 3$ modes in addition to the two described above.  However, these other $\ell = 3$ Fuller modes have components split by either 2 or 4 times the orbital frequency and can be
difficult to distinguish from $Y_{11\pm}$ or $Y_{22\pm}$ modes.

In order to make concrete some of the properties of the new stationary perturbed modes, we have developed a code to simulate such pulsations in a binary star system. We utilize two stationary spherical stars separated by the semi-major axis of the binary $a$. We uniformly distribute 7000 unit vectors on each star, and then view the system from the perspective of an observer orbiting around the binary at frequency $\nu_{\rm orb}$, at an inclination angle $i$. Any pulsation mode can then be assigned to the fixed pulsating star; this mode can be either circulating or stationary\footnote{We assume that the fractional flux, $\delta F/F$, emerging from each surface element of the star due to the pulsation is simply proportional to the $Y_{\rm \ell,m}$ of the particular mode being simulated.}. The pulsation amplitudes are arbitrary since we have no way of calculating these a priori.  At a given time, we evaluate the dot product between the viewing direction's unit vector and the unit vectors over the surface of both stars. If the dot product is negative, then the observer sees that part of the star; all the visible flux (as calculated in this manner) is then summed. Eclipsed regions are also excluded from the sum.  This same dot product is also used to compute the limb darkening and the projected area of the surface element. We also add a simple cosine term of $2 \nu_{\rm orb}$ to represent an ellipsoidal light variation for aesthetic purposes.

This set of calculations is done every 120\,s (to match the TESS 2-min cadence) for 50,000 steps (to roughly match three sectors of TESS data). The simulated time series is cleaned of orbital harmonics, as is done for the real data set. We then take the Fourier transform of the simulated data. The entire process is then repeated for 18 orbital inclination angles between 90$^\circ$ and 5$^\circ$.  An illustrative set of simulated light curves for a $Y_{33+}$ mode is shown in Fig.~\ref{fig:sim_LCs}.  Each light curve has six pulsational maxima per orbit with minima (in this particular case) at both eclipses.  In general, the relative pulsation amplitude decreases with decreasing inclination angle, going to zero at $i = 0$, as expected.  We highlight the curve at $i=75^\circ$ because that is close to the inclination determined for TIC 287869463 (see Sect.~\ref{sec:system}).

Fourier transforms of simulated data for four different $\ell=3$ modes are shown in Fig.~\ref{fig:sim_FTs}. In the upper left panel, we display the FTs for a $Y_{33+}$ mode of the type detected for TIC 287869463 as a function of orbital inclination angle. There are two main peaks that are separated by 6 $\nu_{\rm orb}$.  The smaller peaks for $i \gtrsim 80^\circ$ are due to the eclipses periodically modulating the intensity of the pulsations, a phenomenon known as ``spatial filtration'' (\citealt{gamarova03}; \citealt{reed05}; \citealt{biro11}; \citealt{johnston23}; \citealt{vanreeth23}).  For an inclination of $\sim$75$^\circ$ (as observed for TIC 287869463), these frequency peaks would not have been detected.  In the upper right panel, we show the corresponding FTs for a $Y_{33-}$ mode, and the results are similar to those for a $Y_{33+}$ mode.  The main difference is that the small intermediate peaks are somewhat more pronounced.  This is due to the fact that the pulsation amplitudes are a maximum at the eclipses, rather than a minimum, and this enhances the effects of the eclipses on the pulsation's visibility.  But, even at $i \simeq 75^\circ$, these smaller peaks would have been only barely detectable.

\begin{figure}
\centering
\includegraphics[width=0.48\textwidth]{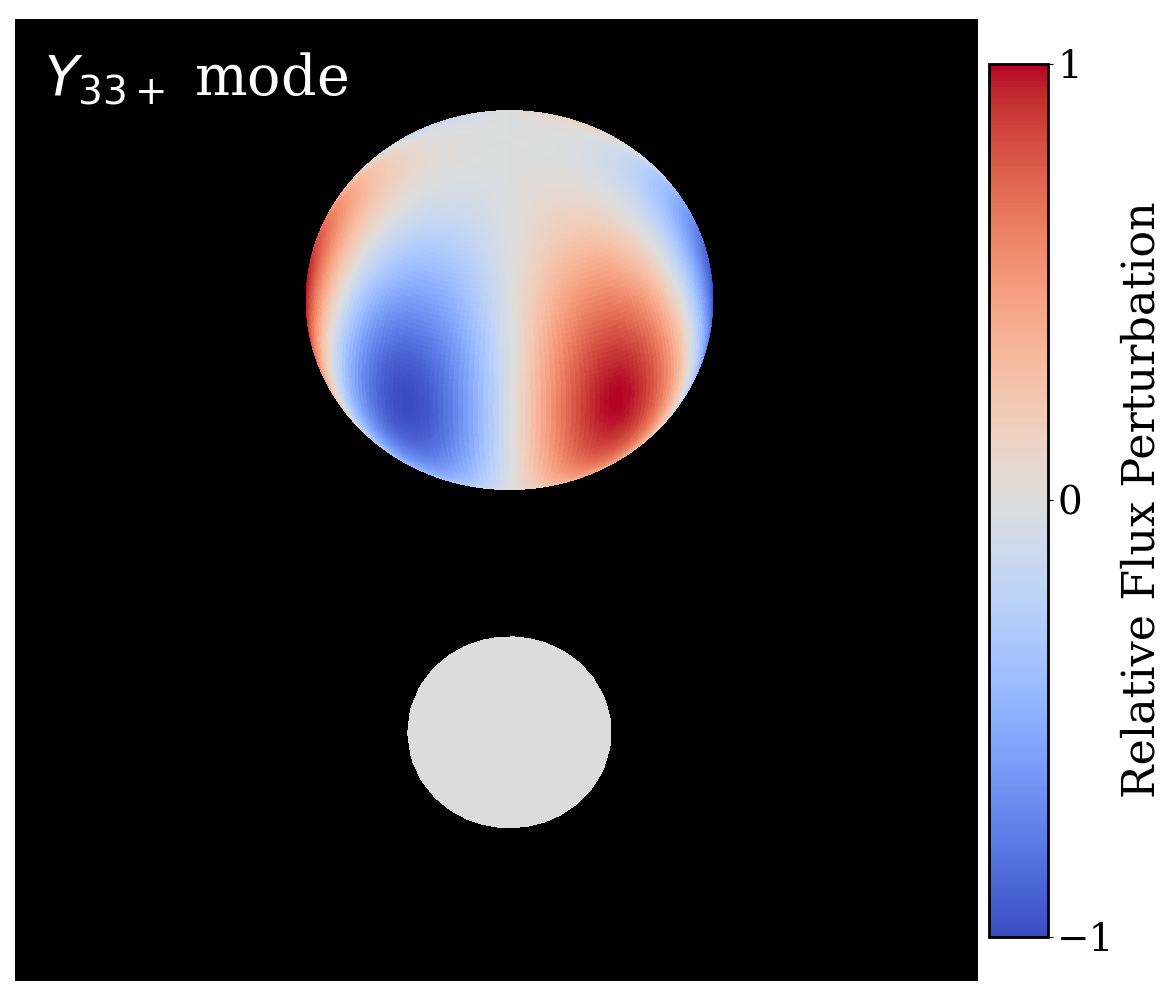}
\includegraphics[width=0.48\textwidth]{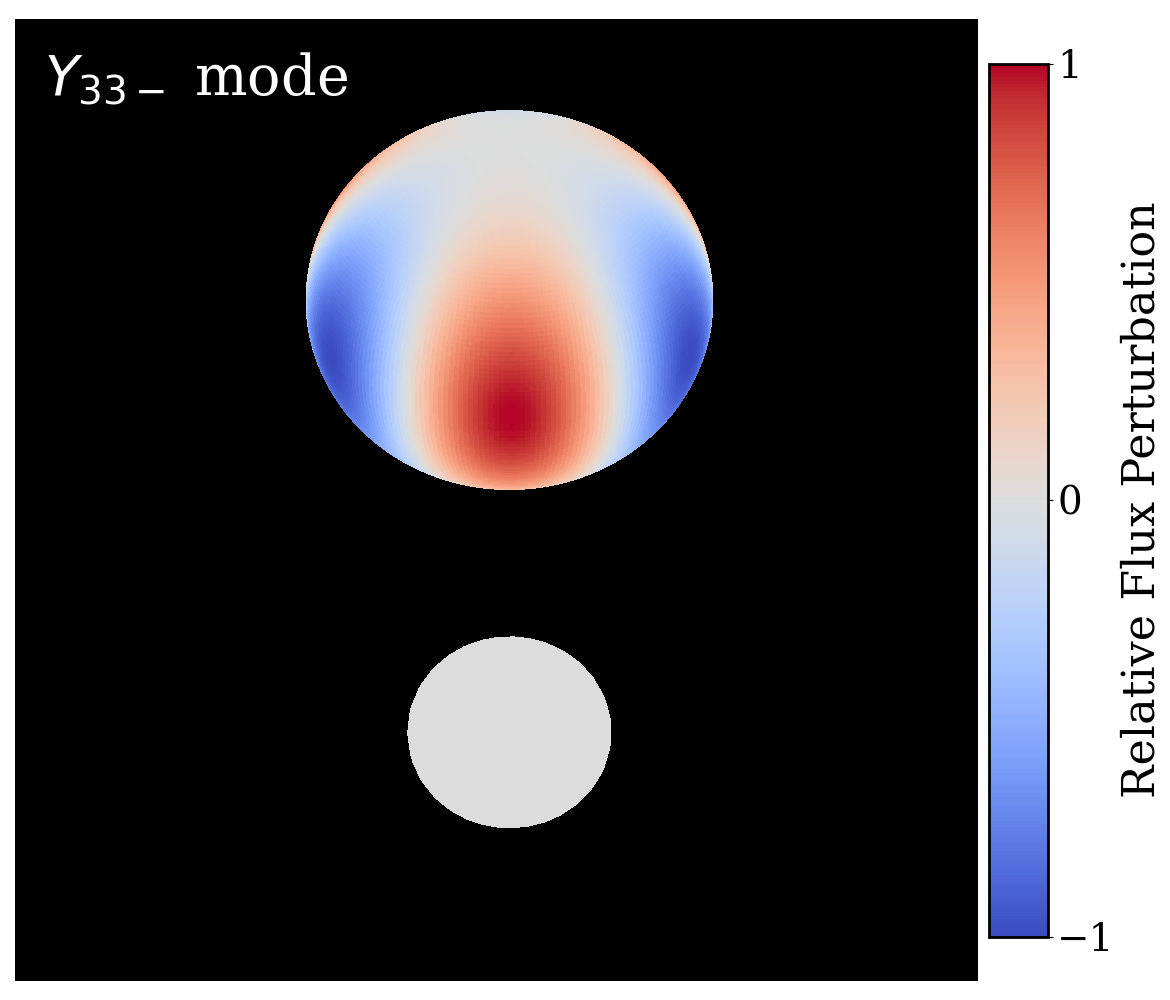}
\caption{Diagrams showing the flux perturbation across the surface of a star pulsating in the $Y_{33+}$ mode (top) and $Y_{33-}$ mode (bottom). The diagram corresponds to orbital phase zero for an observer viewing at at inclination $i=55^{\circ}$ relative to the orbital axis. The observed phase of the amplitude modulation of the O1 mode is consistent with that of a $Y_{33+}$ mode.}
\label{fig:Y33ModeCartoon}
\end{figure}

A pictorial representation of the $Y_{33+}$ mode and the $Y_{33-}$ mode, as they might appear in the context of TIC 287869463, is shown in Figure \ref{fig:Y33ModeCartoon}.

In the lower two panels of Fig ~\ref{fig:sim_FTs} we show the FTs for simulated data of two different $\ell = 3$ modes, namely a $Y_{33x}$ and $Y_{31x}$ mode.  Here, the ``$x$'' signifies that the pulsation axis is tilted into the orbital plane and lies along the tidal axis.  As mentioned above, these are ``tidally tilted'' pulsations, and they have traveling waves propagating around the $x$ axis. As shown by the simulations, there are numerous intermediate peaks out to $\pm 3 \, \nu_{\rm orb}$.  We remind the reader that the central peak of each multiplet is the only pulsation mode frequency; all of the other components of the multiplet are Fourier descriptions of the observed pulsation amplitude and phase modulating with changing orbital aspect.  These {multiplets in the lower two panels} look nothing like what we see in TIC 287869463.  We have simulated all combinations of traveling and stationary modes around all three axes, $x$, $y$, and $z$ with $\ell = 3$, and the only mode that fits the observational data (FT and the phasing of amplitudes and pulse phases with respect to the eclipses) for TIC 287869463 is a $Y_{33+}$ octupole Fuller mode.

  % Figure 8

%%%%%%%%%%%%%%%%%%%%%%%%%%%%%%%%%%%%%%%%%%%%%%%%%%%%%%%%%%%%%%
\section{Analysis of the binary system}
\label{sec:system}
To evaluate the properties of the two stars in TIC 287869463, we simultaneously fit the spectral energy distribution (SED) curve and the TESS light curve. We employ a Markov Chain Monte Carlo (MCMC) algorithm to fit for the stellar masses $M_1$ and $M_2$ (where $M_1$ is the more massive, pulsating star), the system age, the orbital inclination angle $i$, interstellar extinction, $A_V$, and the distance. The latter two parameters could be fixed at the values from Gaia. In total, there are between 4 and 6 free parameters. We assume that the binary has evolved in a coeval fashion, i.e., where there has been no prior mass transfer or mass loss from the system. We thus can use stellar evolution models \citep{choi16,dotter16} to determine the radii and $T_{\rm eff}$ of both stars, which are uniquely determined given their masses and age.  

We collected the available archival spectral flux measurements (SED points) from 0.15 to 11.6\,$\mu$m.  These data were accessed from VizieR\footnote{\url{http://vizier.cds.unistra.fr/vizier/sed/}} \citep{ochsenbein00} which, in turn, utilizes systematic sky coverage of such surveys such as Pan-STARRS \citep{chambers16}, SDSS \citep{gunn98}, 2MASS \citep{2MASS}, WISE \citep{WISE}, GALEX \citep{bianchi17}, and Gaia \citep{GaiaEDR3}.  More details about this type of SED fitting for binary star systems are given in \citet{jayaraman24,handler25,yakut25a,yakut25b}. 

The input information we use includes: (i) 23 available SED points spanning 0.15 $\mu$m to 11.6 $\mu$m, (ii) an approximate value for $K_1$ of the radial velocity (RV) curve\footnote{We utilize the Gaia parameter RV$\_$amp$\_$robust of 144 km s$^{-1}$ as a proxy for twice the RV amplitude $K_1$ (e.g., \citealt{katz23}). }, (iii) the eclipsing light curve profile, (iv) the Gaia distance of $1086 \pm 3$ pc \citep{bailer-jones21}, and a Gaia value for $A_G$ of 0.709\footnote{This translates to an $A_V$ of $\sim$0.84 using either the rule-of-thumb factor of 1.2 or the detailed expression of \citet{danielski18}; in particular their equation (3) and coefficients from row 5 of their Table 2.}.  Additionally, we use the MIST evolution tracks \citep{paxton11,paxton15,choi16,dotter16,paxton19} to relate the stellar mass and age to its radius and $T_{\rm eff}$.  In order to compute the model stellar spectra from the radii and $T_{\rm eff}$ of the stars, we make use of the Castelli \& Kurucz model stellar atmospheres \citep{castelli03}, assuming a solar metallicity.
Finally, we correct for interstellar extinction using the prescription presented in \citet{cardelli89}.

When fitting a composite SED of two or more stars, it is important to have additional constraints between the properties of one star and the other(s).  In particular, the eclipsing light curve in this system provides important information about such things as $T_{\rm eff ,2}/T_{\rm eff ,1}$, $R_2/R_1$, $R_1/a$, and the orbital inclination angle, where $a$ is the semi-major axis of the system.  Since well-developed light curve emulators such as {\sc Phoebe} \citep{prsa05} are not designed to work jointly with SED fitting codes, we have incorporated our own simple light curve emulator to help the SED fitter more properly separate the light contribution from the two stars \citep{jayaraman24}. 
This light curve emulator utilizes two spherical stars with limb darkening to generate eclipses.  The out-of-eclipse behavior is described by three sinusoids of the form $\cos(\omega_{\rm orb} t)$, $\cos(2 \omega_{\rm orb} t)$, and $\sin(\omega_{\rm orb} t)$ to represent such effects as ellipsoidal light variations, illumination effects, Doppler boosting, and corotating spots on the cooler star (see, e.g., \citealt{kopal59,carter11}).

\begin{figure}
\centering
\includegraphics[width=0.98\columnwidth]{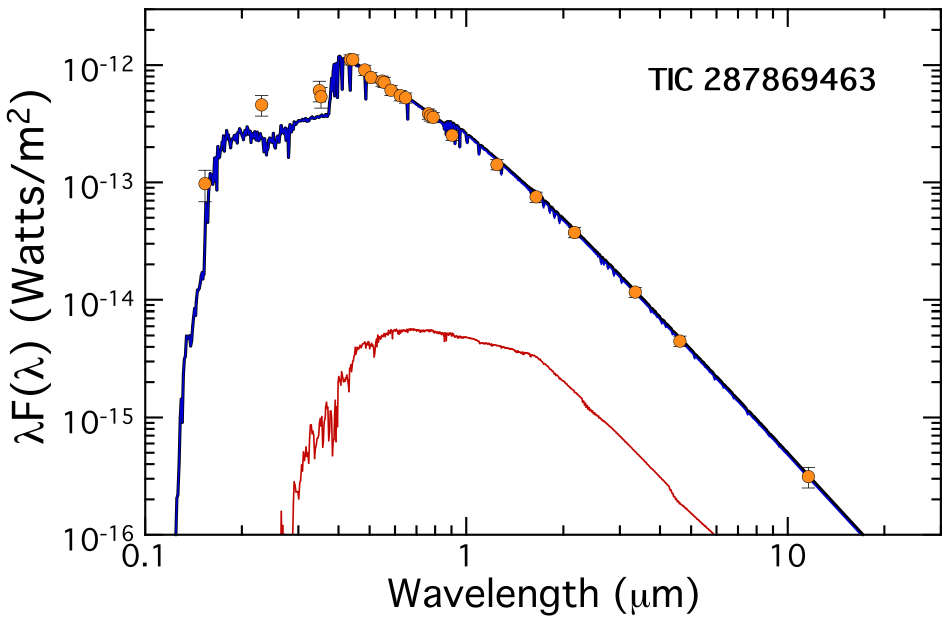}
\includegraphics[width=0.98\columnwidth]{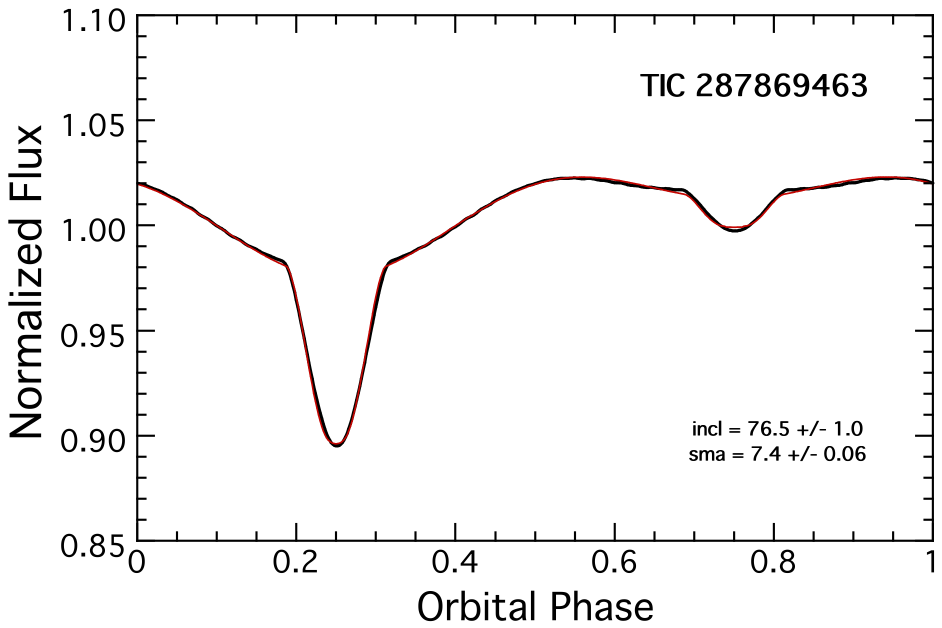}
\caption{Top panel: SED fit for TIC 287869463. The blue curve is the model spectrum for the primary star (the pulsator), and the red curve represents the model spectrum for the cooler, smaller secondary star.  The black curve, nearly coinciding with the red curve, is the sum of the model fluxes.  The orange points are measured SED points from the literature.  For the SED fitting techniques and the origin of the data points, see Sect.~\ref{sec:system}. \textit{Bottom panel}: Light curve fit using our custom light curve emulator.  The black curve is the Fourier reconstructed light curve (i.e., the `data') and the red curve is the model fit.  See text for details.}
\label{fig:SED}
\end{figure}  % Figure 9

The best-fitting SED and light curve models are shown in the top and bottom panels of Fig.~\ref{fig:SED}, respectively.  The best-fitting system parameters derived from these fits are summarized in Table \ref{tbl:system_parms}.  The primary star, with a mass of 2.2 M$_\odot$, radius 2.6 R$_\odot$, and $T_{\rm eff,1} = 8734$ K, has a luminosity typical of a $\delta$ Scuti star, with $T_{\rm eff}$ near the blue border of the $\delta$~Sct instability strip.  The pulsator's parameters place it between the \citet{balona15} $\delta$-Scuti-category boxes 4 and 7 in their Fig.~2.  Pulsators in these categories can have pulsations up to $\sim$50 d$^{-1}$, and there are many such systems with pulsations in a range similar to that which we see in TIC 287869463.

Finally, we point out that even though the secondary star contributes less than 1\% of the system luminosity, we can be confident of having measured its parameters.  The reason is that there are numerous constraints between the secondary and the primary other than from the SED, e.g., $K_1$, $R_2/R_1$, and $T_2/T_1$.  Therefore, the primary star largely accounts for matching the SED, while the properties of the secondary star (including its contribution to the SED) come from the relative relations listed above, and a common age.

\begin{figure}[h]
\centering
\includegraphics[width=0.98\columnwidth]{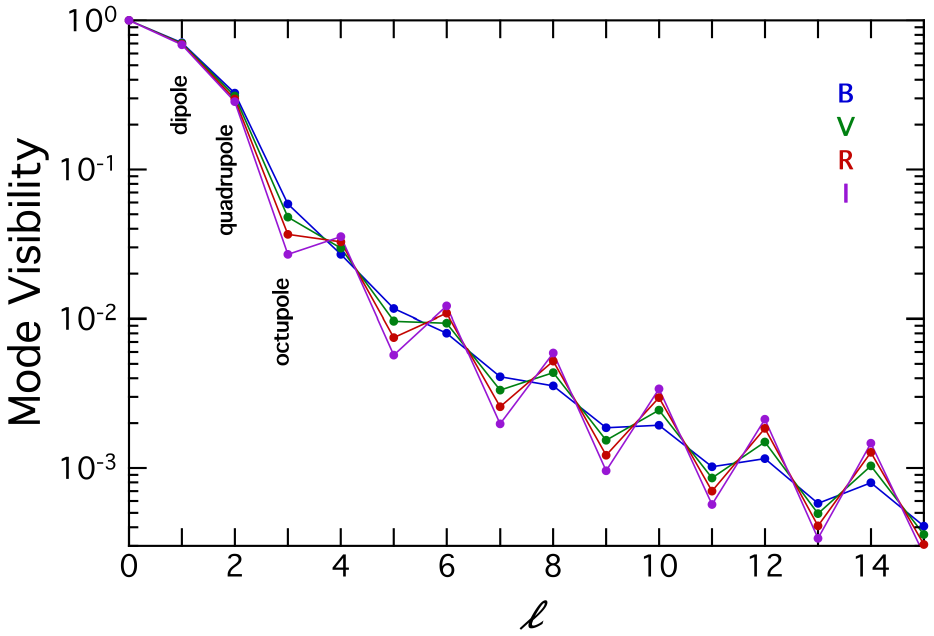}
\caption{Mode visibility diagram as a function of $\ell$. These have been calculated as in \citet{daszynska02}, for $BVRI$ passbands; the red passband most closely matches that of TESS.}
\label{fig:visibility}
\end{figure}  % Figure 10

\begin{table}
\centering
\caption{Binary system parameters$^a$}
\small
\begin{tabular}{lc}
\hline
\hline
Parameter & Value \\
\hline
$M_1$ [M$_\odot$] &  $2.18 \pm 0.05$ \\
$M_2$ [M$_\odot$]  &  $0.80 \pm 0.03$ \\ 
$R_1$ [R$_\odot$]  & $2.57 \pm 0.07$  \\
$R_2$ [R$_\odot$]  & $0.73 \pm 0.03$  \\
$T_{\rm eff,1}$ [K] & $8734 \pm 200$  \\
$T_{\rm eff,2}$ [K] &  $4943 \pm 150$  \\
$L_1$ [L$_\odot$] &  $34.7 \pm 3.7$ \\
$L_2$ [L$_\odot$] &  $0.29 \pm 0.04$\\
a [R$_\odot$] & $7.5 \pm 0.3$  \\
Orbital Inclination [$^\circ$] & $76.5 \pm 1.0$ \\
$P_{\rm orb}$ [d]$^b$ & 1.374\,530 \\
Primary Eclipse [BJD]$^b$ & 2460014.476 \\
$K_1$ [km s$^{-1}$]$^c$ & $\sim$72\\
$A_V$ & $0.80 \pm 0.07$\\
$A_G$$^d$ & 0.709 \\
Distance [pc]$^e$ & $1086 \pm 3$ \\
age [Myr] & $630 \pm 50$ \\
\hline
\label{tbl:system_parms}  % Table 4
\end{tabular}

\textit{Notes.}  (a) Based on the SED and light curve fitting, as described in Sect.~\ref{sec:system}, unless otherwise noted. (b) This work. (c) Half the value of the Gaia parameter RV\_amp\_robust \citep{katz23}.  (d) Gaia's extinction.  (e) Gaia's distance \citep{bailer-jones21}.

\end{table}

%%%%%%%%%%%%%%%%%%%%%%%%%%%%%%%%%%%%%%%%%%%%%%%%%%%%%%%%%%%%%%

\section{Discussion}
\label{sec:discussion}

While carrying out a large scale search for Fuller-mode pulsators in TESS data, we discovered an unprecedented octupole mode in a $\delta$ Scuti star.  It is characterized by two peaks in the Fourier transform separated by $6 \, \nu_{\rm orb}$.  This splitting corresponds to 6 times the orbital frequency to within an accuracy of one part in $10^5$.   We argue that this cannot be just two random and unrelated modes of the pulsating star, but must comprise a single mode. 

This octupole mode is actually a combination of ordinary $Y_{3+3}$ and $Y_{3-3}$ modes that are aligned with the stellar rotation axis.  These are combined into a new eigenmode of the star by binary perturbations, and we call it a Fuller $Y_{33+}$ mode \citep{fuller25}.  We presented a number of compelling reasons why the octupole mode cannot plausibly be two independent modes whose frequencies have an accidental alignment in the echelle diagram.  This is the first octupole mode securely identified in any $\delta$-Scuti star, and the first Fuller-type $\ell = 3$ stationary {sectoral} mode found in any star, including the Sun.

As we show in the Appendix, the pulsation frequencies of both the dipole and octupole modes are found to be steadily increasing with time, all at different rates.  However, the splits in the mode frequencies remain constant at an integer multiple of the orbital frequency.

Because of geometric cancellation of higher-order $\ell$ modes over the visible hemisphere of pulsating stars, there is sharp decrease in the ``visibility'' of a pulsation mode with increasing $\ell$.   To investigate the visibility of higher degree modes we follow the approach described in \citet{daszynska02}. As a first step, we approximate the mode visibility by the disc-averaging factor 
\begin{equation} 
b_\ell = \int_0^1 h(\upmu) P_\ell(\upmu) \upmu d\upmu
\end{equation}
where $\upmu$ is the cosine of the angle between the observer's line of sight and the local surface normal, and $h$ is the limb darkening function.  This should be reasonably accurate if the effects of rapid rotation and tidal interaction are neglected.  In a more general approach, as in the case of the star we studied, the mode visibility may also depend on the azimuthal order $m$.

We show in Fig.\,\ref{fig:visibility} a plot of the visibility of pulsation modes in a star like TIC 287869463 as a function of $\ell$ for four different filters, including a red filter which is a reasonable approximation to the TESS bandpass (the results are also given numerically in Table \ref{tbl:visibility}). Dipole and quadrupole modes ($\ell =1$ and $\ell =2$, respectively) have relatively high visibilities of 0.7 and 0.3, respectively.  However, for octupole modes ($\ell=3$) the visibility drops abruptly to 0.037 (for the red filter). In TIC 287869463, the dipole modes have amplitudes of 1.8 mmag and 0.8 mmag.  However, the amplitude of the octupole mode is not too far below those, at 0.4 mmag.  If we renormalize this according to the relative visibilities, we find that the octupole mode has an intrinsic amplitude of $\sim$10\,mmag. Pulsation amplitudes of this size are fairly common among $\delta$ Scuti stars. 

This exercise also shows that even hexadecapole modes with $\ell = 4$ should be as straightforward to detect as the octupole mode found here. However, higher modes with $\ell \ge 5$ will be very difficult to detect.

It is also worth noting that according to the simulations of \citet{daszynska06}, the octupole mode should be relatively easy to identify from radial-velocity variations. Therefore, time-series spectroscopy could be used to validate our octupole mode identification for TIC 287869463.

\begin{figure}[h]
\centering
\includegraphics[width=0.98\columnwidth]{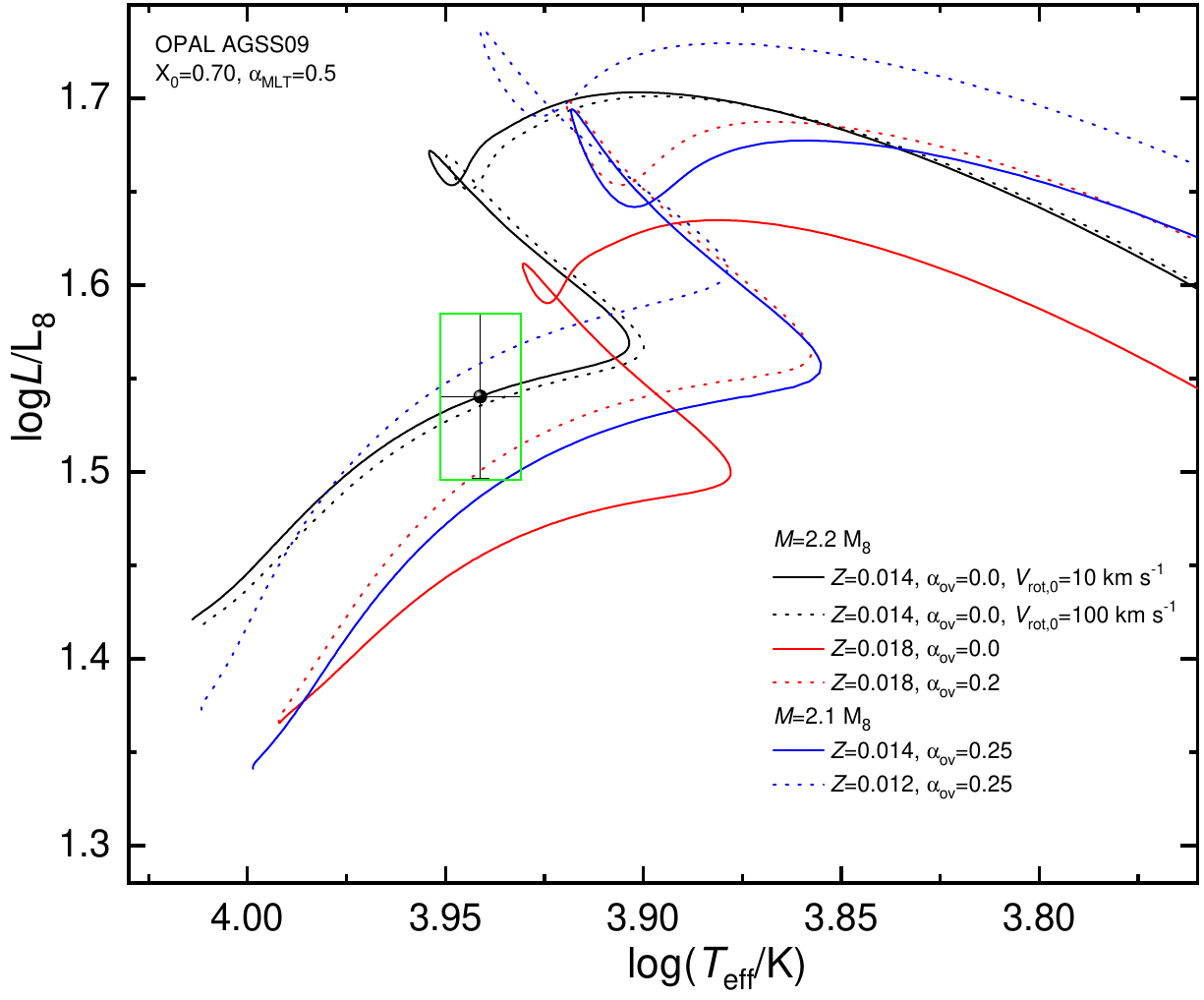}
\caption{Six different evolution tracks in the HR diagram compared to the measured luminosity and $T_{\rm eff}$ of TIC 287869463.  The parameters that are varied in the different tracks involve metallicity $Z$, $\alpha_{\rm MLT}$, mass, rotation velocity, and convective overshooting, $\alpha_{\rm ov}$ (see text for details).}
\label{fig:HR}
\end{figure}

Finally, we show in Figure \ref{fig:HR} model evolution tracks in the HR diagram for several sets of model parameters, along with a comparison to the stellar values we derived for the TIC 287869463 pulsator.  Evolutionary models were computed with the Warsaw-New Jersey code (e.g., \citealt{pamyatnykh99}) adopting the OPAL opacity tables \citep{iglesias96} and chemical mixture of \citet{asplund09}.  Given the accurately determined stellar parameters, including the mass and radius, one may expect strong constraints from seismic modeling of this star, for example on the metallicity $Z$, the efficiency of outer-layer convection $\alpha_{\rm MLT}$, and the overshooting distance $\alpha_{\rm ov}$. This will be the subject of our follow-up studies of this object.

  % Figure 11

\begin{table}
\centering
\caption{Mode Visibilities$^a$}
\small
\begin{tabular}{lcccc}
\hline
\hline
$\ell$ & B & V & R & I \\
\hline
1 & 0.70930 & 0.70160	&  0.69360  &  0.68660 \\
2 & 0.32490  & 0.31130	& 0.29710 &  0.28480 \\
3 & 0.05868 & 0.04797	& 0.03668 &  0.02690 \\
4 & 0.02688 &	0.02969	& 0.03267 & 0.03535 \\
5 & 0.01167 &	0.00964	& 0.00750  & 0.00572 \\
6 & 0.00802 & 0.00937	& 0.01085 & 0.01214 \\
7 & 0.00410 &	0.00333	& 0.00258 & 0.00198 \\
8 & 0.00351 &	0.00436	& 0.00521 & 0.00592 \\
9 & 0.00186 &	0.00154	& 0.00121 & 0.00096  \\
10 & 0.00194	& 0.00245 & 0.00297 & 0.00340 \\
11 & 0.00102	& 0.00086 & 0.00070 & 0.00057 \\
12 & 0.00116	& 0.00149 & 0.00184 & 0.00212 \\
13 & 0.00058  & 0.00049  & 0.00041 & 0.00034 \\
14 & 0.00080  & 0.00103  & 0.00127 & 0.00147 \\
15 & 0.00041  & 0.00036   & 0.00031 & 0.00026 \\
\hline
\label{tbl:visibility}  % Table 5
\end{tabular}

\textit{Notes.}  (a) Computed for the $B$, $V$, $R$, and $I$ wavebands.  The same values used to generate Figure \ref{fig:visibility}. 

\end{table}

%%%%%%%%%%%%%%%%%%%%%%%%%%%%%%%%%%%%%%%%%%%%%%%%%%%%%%%%%%%%%%

%%%%%%%%%%%%%%%%%%%%%%%%%%%%%%%%%%%%%%%%%%%%%%%%%%%%%%%%%%%%%%
\begin{acknowledgements}
RJ is currently supported by a Klarman Fellowship from the College of Arts \& Sciences at Cornell University.
This project has received partial funding from the HUN-REN Hungarian Research 
Network. T.B. acknowledges the financial support of the Hungarian National 
Research, Development and Innovation Office -- NKFIH Grant OTKA K-147131.
VBK is grateful for financial support from NASA grant 80NSSC22K0747. GH thanks the Polish National Center for Science (NCN) for support through grant 2021/43/B/ST9/02972.
JDD thanks the Polish National Center for Science (NCN) for support through grant 2023/50/A/ST9/00144.
This paper includes data collected by the TESS mission. Funding for the TESS mission is provided by the NASA Science Mission Directorate. The QLP data used in this work were obtained from MAST \citep{huang20}, hosted by the Space Telescope Science Institute (STScI). STScI is operated by the Association of Universities for Research in Astronomy, Inc., under NASA contract NAS 5-26555. This work also presents results from the European Space Agency (ESA) space mission Gaia. Gaia data are being processed by the Gaia Data Processing and Analysis Consortium (DPAC). Funding for the DPAC is provided by national institutions, in particular the institutions participating in the Gaia Multilateral Agreement. The Gaia mission website is https://www.cosmos.esa.int/web/gaia. The Gaia archive website is https://gea.esac.esa.int/archive/. 
Some of the SED fluxes and magnitudes were obtained with the
Wide-field Infrared Survey Explorer, which is a joint project of the University of 
California, Los Angeles, and the Jet Propulsion Laboratory/California Institute 
of Technology, funded by the National Aeronautics and Space Administration. 
Additionally, some of the SED fluxes and magnitudes were obtained with the 
Two Micron All Sky Survey, which is a joint project of the University of Massachusetts 
and the Infrared Processing and Analysis Center/California Institute 
of Technology, funded by the National Aeronautics and Space Administration 
and the National Science Foundation. We used the Simbad service operated by 
the Centre des Donn\'es Stellaires (Strasbourg, France). This research has also 
made use of the VizieR catalogue access tool, CDS, Strasbourg, France (DOI 
: 10.26093/cds/vizier). The original description of the VizieR service was published in \citet{ochsenbein00}.
\end{acknowledgements}

%%%%%%%%%%%%%%%%%%%%%%%%%%%%%%%%%%%%%%%%%%%%%%%%%%%%%%%%%%%%%%

% Please note that we have included the references below in
% order to compile the document, but we ask you to:

%%%%%%%%%%%%%%%%%%%%%%%%%%%%%%%%%%%%%%%%%%%%%%%%%%%%%%%%%%%%%%%
% Appendices must be placed after   \end{thebibliography}
% They will be placed automatically on a new page.
%%%%%%%%%%%%%%%%%%%%%%%%%%%%%%%%%%%%%%%%%%%%%%%%%%%%%%%%%%%%%%%
\begin{appendix}
%%%%%%%%%%%%%%%%%%%%%%%%%%%%%%%%%%%%%%%%%%%%%%%%%%%%%%%%%%%%%%%
% In the PDF output, floats should be placed
% under their own appendix, not before the title, nor after the
% title of the next appendix.

% In short appendices, onecolumn floats (\figure*
% or \table*) will generate a blank page.
% To prevent this behaviour, a few examples are provided here. 

% In case you have a lot of floating objects for little text and the 
% LaTeX engine moves the floats away from their context, the command
% \FloatBarrier of the “placeins” package will empty the
% float buffer and place all stored floats in the continuity.

% If you still encounter problems with wide floats placement,
% just use the onecolumn environment throughout the appendices.
%%%%%%%%%%%%%%%%%%%%%%%%%%%%%%%%%%%%%%%%%%%%%%%%%%%%%%%%%%%%%%%

%____________________________________________________________
%       Wide floats at the start of an appendix: first method
%-------------------------------------------------------------
% To prevent a blank page after the start of an appendix:
% - Switch to one \onecolumn first
% - Declare the section title
% - Declare the onecolumn float with the parameter [ht!]
% - Revert to \twocolumn at the end of the section
\onecolumn
\section{Changes in the pulsation frequencies and the orbital period}
\label{sec:phase_tracking}
%
%In the PDF output, \underline{floats should be placed
%under their own appendix}, not before the title, nor after the
%title of the next appendix. In short appendices, one-column floats
%\{figure*\} or \{table*\} will generate
%a blank page. To prevent this behaviour, we recommend to switch
%to \textbackslash onecolumn and set the [ht!] parameter 
%in your floats: please check the \LaTeX code of this appendix.\\

%In case you have a lot of floating objects for little text and the 
%\LaTeX engine moves the floats away from their context, the command
%\textbackslash FloatBarrier of the “placeins” package will empty the
%float buffer and place all stored floats in the continuity. If you still encounter problems with wide floats placement, just use the \textbackslash onecolumn
%environment throughout the appendices.

Both the pulsation frequencies and the orbital period appear nearly constant across the 200-s cadence data from the four contiguous TESS sectors 63-66, or even to some extent when we include the final three sectors of 200-s cadence---separated from the former sectors by about 700\,d.  However, the apparently constant pulsation frequencies are seen to be clearly varying once we add five earlier sectors with 600-s cadence data (sectors between S30 and S39). Overall, after analyzing the available TESS data between sectors S30 and S93, the pulsation frequencies are all found to be significantly increasing with time.  The orbital frequency can be tracked to even earlier times by using six sectors between S3 and S13 with 30-min cadence, and $\nu_{\rm orb}$ is also found to be increasing with time.  We thus sought to characterize the frequency changes in the pulsations as well as in the orbit.

To track the pulsation phases $\phi$ (see also \citealt{bowman16}), we fit a sine curve of the form $A \cos\left[\omega (t -t_{\rm ecl})- \phi\right]$ to represent a pulsation over an interval of, say, half of a full TESS sector ($\sim$12\,d).  We needed to consider sufficiently long intervals because some of the pulsation peaks are within a fraction of 1 d$^{-1}$ of other peaks.  The reference time, $t_{\rm ecl}$ is the time of a primary eclipse, e.g., BJD 2460014.476.  We then fit for $A$ and $\phi$ and stepped the fitting window forward in time by a fraction of a TESS sector.  

\begin{figure}
\centering
\includegraphics[width=0.48\columnwidth]{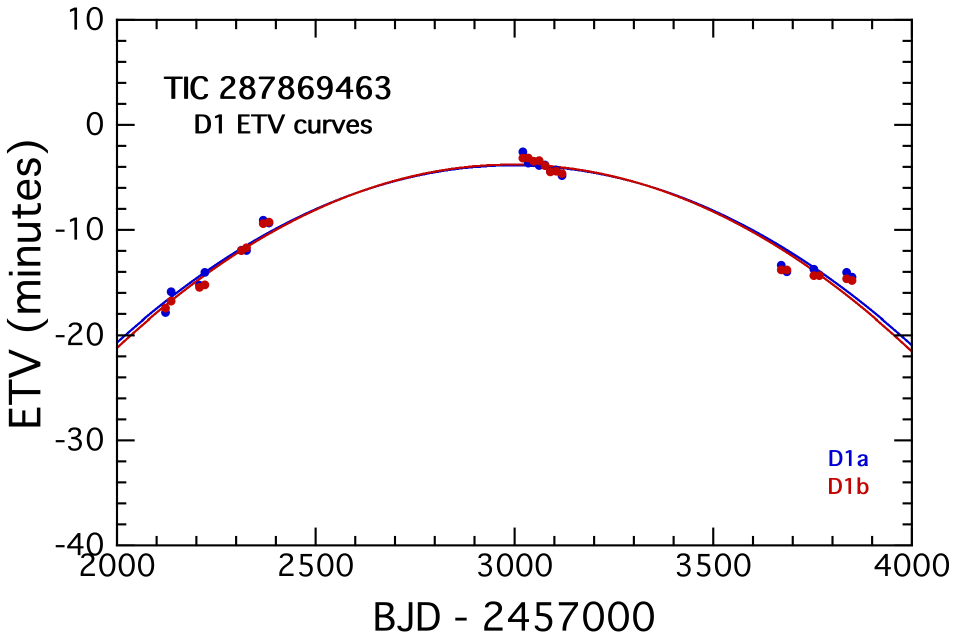}
\includegraphics[width=0.48\columnwidth]{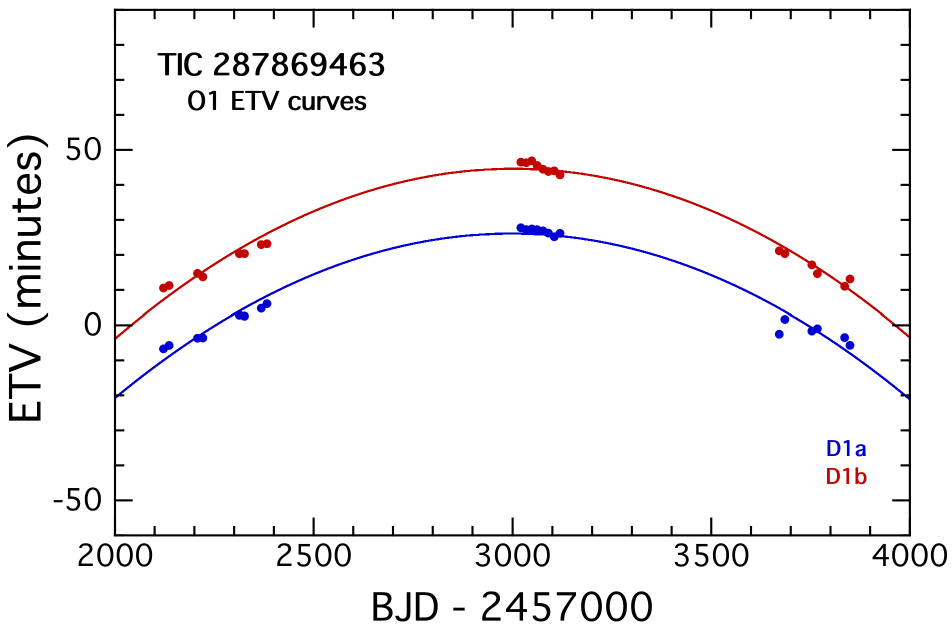}
\includegraphics[width=0.48\columnwidth]{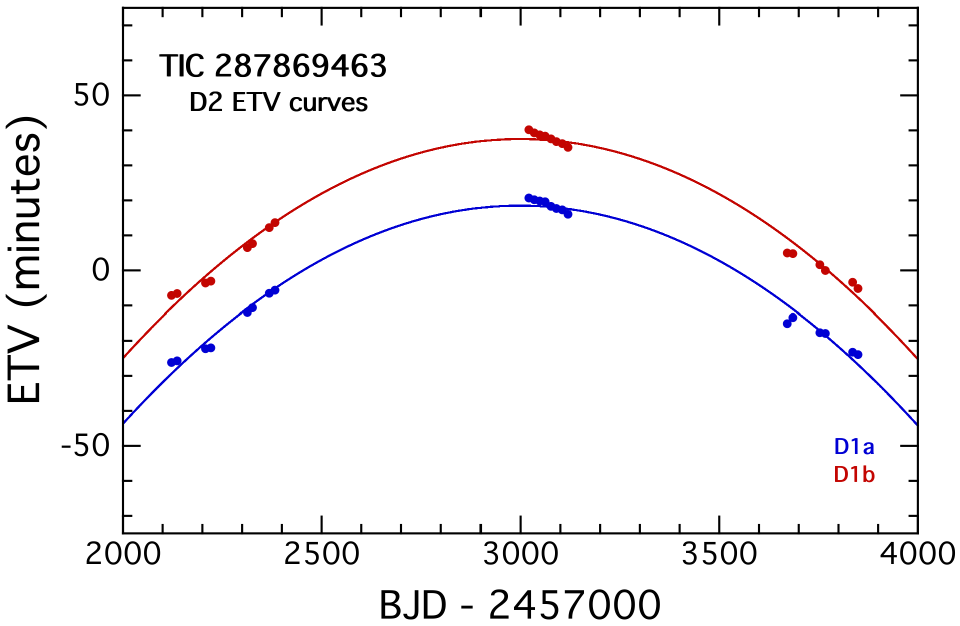}
\includegraphics[width=0.48\columnwidth]{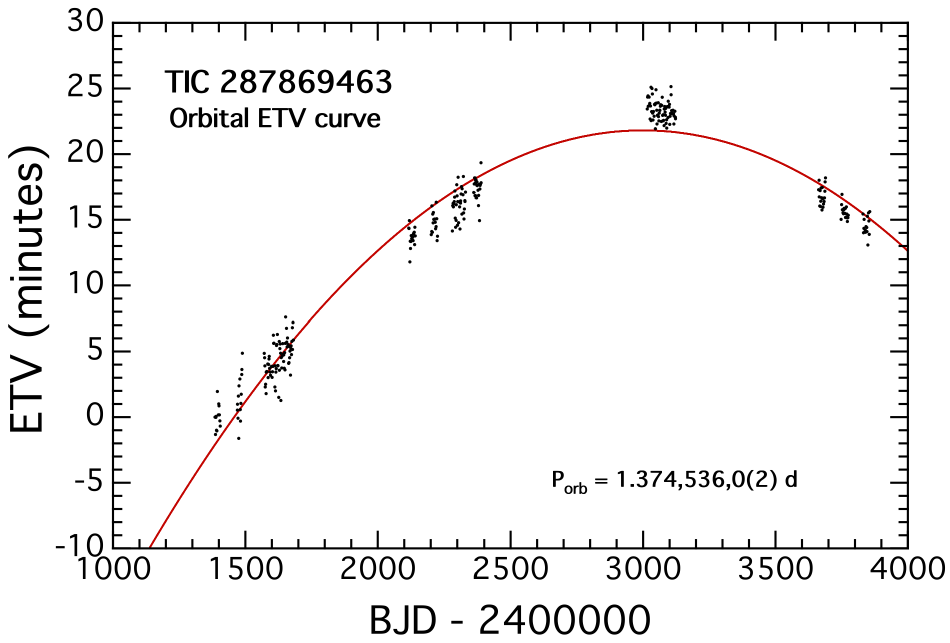}
\caption{ETV curves for each component of the three pulsation modes, D1, D2, and O1, as well as from the binary orbital period.  All four curves show decreasing periods with time, but not at the same rates.  All phases are referenced to the times of a primary eclipse, i.e., at BJD - 2,456,014.4757. Note that the vertical scale for the D1 ETVs is exactly 1/3 that for the D2 and O1 ETVs. In turn, the vertical scale for the orbital ETV curve is 4/5 that of the D1 ETVs.}
\label{fig:etvs}
\end{figure}

The results of tracking the phase for each of the 6 pulsation frequencies are given in Figure \ref{fig:etvs}.  Each panel displays the equivalent Eclipse Timing Variation (ETV) curve for both components of a given pulsation mode: D1a and D1b, D2a and D2b (top two panels) and O1a and O1b in the lower left panel.  Note  that for mode D1, the $a$ and $b$ components track each other very well, and we saw in Fig.~\ref{fig:amp_phase} that indeed these two components are always in phase at the primary eclipses.  By contrast, we can see from the ETVs of D2 and O1 that the ETV curves for the two components are separated by close to half a pulsation period, consistent with the fact that the two components of both modes remain out of phase at the primary eclipses over all the sectors with 10-min or better cadence.  

The final panel in Fig.~\ref{fig:etvs} shows the ETV curve for the orbital period. This shows that $P_{\rm orb}$ is also apparently decreasing with time.  If we interpret this as an orbital decay, then the decay timescale of just $\sim$$2 \times 10^5$\,yr seems rather implausible.  Alternatively, we might interpret the curve as a portion of a Doppler delay orbit caused by an orbiting third body.  Unfortunately, the duration of available TESS data is insufficient to determine an outer orbit for a putative third body.  However, for one of the shorter allowed orbits of 15\,yr, the semi-major axis is 950 light-seconds (for an assumed circular orbit), and the corresponding mass function is $f(M) \simeq 0.03$\,M$_\odot$. For a binary mass of 3\,M$_\odot$ (see Table \ref{tbl:system_parms}), this yields a value of $M_3 \sin^3 i $ of about 0.75\,M$_\odot$.  A third body of such a mass would not have contributed sufficient light to the SED to have been unequivocally detected, especially for outer inclination angles of $\gtrsim 75^\circ$.  We note that the known secondary in the binary contributes only $\lesssim 0.01$ of the system luminosity, and a comparable third star would barely perturb the SED. 

\begin{table*}[h]
\centering
\caption{Summary of frequencies and frequency derivatives in TIC 287869463$^a$}
\small
\begin{tabular}{llcc}
\hline
$\nu$ & $\nu_{\rm split}$$^b$ & $\dot{\nu}$$^c$  & diff.~in $\dot{\nu}$  \\
 d$^{-1}$ & d$^{-1}$ & 10$^{-6}$ d$^{-2}$ & 10$^{-8}$d$^{-2}$  \\
\hline
38.45620(1)   & 0.727516(20) $\times 2$ & 0.90(4) & 7(6)  \\  
39.91123(1)   &   0.727516(20) $\times 2$  & 0.97(4) & 7(6) \\
36.36627(2)   &  0.727514(34)\,$\times 2$ & 3.15(8)  & 7(11) \\
37.82130(2)  & 0.727514(34)\,$\times 2$ & 3.28(8) & 7(11) \\
34.94617(3)  & 0.727518(39) $\times 6$ & 2.47(9) & 18(12) \\
39.31127(3)  & 0.727518(39) $\times 6$ & 2.65(9) & 18(12) \\
\hline
\label{tbl:freq_summary}  % Table 6
\end{tabular}

\textit{Notes.}  (a) All frequencies are referenced to an epoch of BJD 2460000.  Values in parentheses are the uncertainties in the last digits. (b) For reference, the orbital frequency at the same epoch is 0.727,518,2(1) d$^{-1}$ or $P_{\rm orb}$ = 1.374,536,0(2). (c) For reference, $\dot{\nu}_{\rm orb} \simeq 0.009$ in the same units. 
\end{table*}

  % Figure 8

Returning to the systematic non-linear behavior in the ETV curves for the pulsations, we examine what the consequences would be for different assumptions about the nature of the non-linear ETV curve observed for the orbit.  Suppose that a Fuller mode pulsation frequency is $\nu_{\rm p}$, and that it is physically changing with time as $\dot{\nu}_{\rm p}$.  The splitting with the orbit will be $\pm m \, \nu_{\rm orb}$ (see, e.g., \citealt{shibahashi12}).  Now, we also allow that the orbital frequency is physically changing as $\dot{\nu}_{\rm orb}$.  This means that the two components of the mode 
(``upper'' and ``lower'') should follow different functions of time:
\begin{eqnarray}
\nu_{\rm p,u} & = & \nu_{\rm p} +m\,\nu_{\rm orb} + \dot{\nu}_{\rm p} t + m\,\dot{\nu}_{\rm orb} t \\
\nu_{\rm p,l} & = & \nu_{\rm p} -m\,\nu_{\rm orb} + \dot{\nu}_{\rm p} t - m\,\dot{\nu}_{\rm orb} t
\end{eqnarray}
In turn, if we were to measure the $\dot{\nu}_{\rm p}$ of the upper and lower components, we should find
\begin{eqnarray}
\dot{\nu}_{\rm p,u} & = &  \dot{\nu}_{\rm p} + m\,\dot{\nu}_{\rm orb}  \\
\dot{\nu}_{\rm p,l} & = &   \dot{\nu}_{\rm p} - m\,\dot{\nu}_{\rm orb} 
\end{eqnarray}
with a difference in $\dot{\nu}_{\rm p}$ of $2 m\,\dot{\nu}_{\rm orb}$. From Table \ref{tbl:freq_summary} we can see that the differences between the $\dot{\nu}$ terms of the upper and lower frequency components are not statistically significant, and the uncertainties are $\sim$10$^{-7}$ d$^{-2}$.  For the octupole mode, we might expect a change in the difference of the $\dot{\nu}$'s by about $6 \times 9 \times 10^{-9} = 5.4 \times 10^{-8}$ d$^{-2}$ due to the changing orbit.  This is smaller than the uncertainty in the difference in $\dot{\nu}$ listed in Table \ref{tbl:freq_summary} for the octupole mode.  Thus, we cannot quite use this effect to test for a decaying orbit.

For Doppler shifts of the binary (which include the pulsating star) the quantity $\dot{\nu}/\nu$ would be constant.  So, we should find that all the pulsation components have $\dot{\nu}_{\rm p}/\nu_{\rm p} = \dot{\nu}_{\rm orb}/\nu_{\rm orb}$, or $\dot{\nu}_{\rm p} \simeq 50 \,\dot{\nu}_{\rm orb}\simeq 0.5 \times 10^{-6}$ d$^{-2}$.  In all cases, this is less than the values of $\dot{\nu}$ for the pulsations listed in Table \ref{tbl:freq_summary}, but would be readily detectable if we could separate the Doppler from other natural changes in pulse frequency.

Regarding the observed (or apparent) changes in the orbital ETV curve, we point out that many such close binaries containing an early-type star exhibit ETVs that have non-linear variations on long timescales (decades or even centuries) that are almost certainly neither actual orbital period changes (i.e., orbital decay) nor third-body induced light travel-time effects.  Some examples of very irregular and unexplained ETVs are the cases of XZ And \citep{yuan19}, X Tri \citep{lee23}, RZ Cas \citep{lehmann20}, and Y Cam \citep{celik24}. While we are confident in the non-linear behavior of the TIC 287869463 orbital ETV curve, we cannot be certain what causes this behavior.  Therefore, at least tentatively, we take all the $\dot{\nu}$'s for the pulsations to be independent of the orbit, and for the orbital $\dot{\nu}_{\rm orb}$ to be due to neither an orbital decay nor an orbital Doppler shift.

\end{appendix}
\end{document}